\documentclass[a4paper,10pt]{article}
\pdfoutput=1 

\usepackage[utf8]{inputenc}

\usepackage{jheppub} 

\usepackage{float}
\usepackage{graphicx}
\usepackage{subcaption}
\usepackage{bbm}
\usepackage{siunitx}
\usepackage{listings}

\usepackage{bbold}

\usepackage{graphicx,epsf}
\usepackage{amsmath,amsfonts,amssymb,amsbsy}
\usepackage{mathtools}

\usepackage[dvipsnames]{xcolor}

\usepackage{hyperref}
\hypersetup{
    colorlinks=true,
    allcolors={blue!60!black}
}

\usepackage{multirow}
\usepackage{here}
\usepackage{stackrel}

\usepackage{ifdraft}
\usepackage[obeyFinal]{easy-todo}

\usepackage{placeins}
\usepackage{slashed} 

\usepackage{booktabs}
\usepackage{adjustbox}

\usepackage{cleveref}

\DeclareMathOperator{\Tr}{Tr}
\DeclareMathOperator{\tr}{tr}

\def\spa#1.#2{\left\langle#1\,#2\right\rangle}
\def\spb#1.#2{\left[#1\,#2\right]}
\def\spash#1.#2{\spa{\smash{#1}}.{\smash{#2}}}
\def\spbsh#1.#2{\spb{\smash{#1}}.{\smash{#2}}}
\def\sand#1.#2.#3{%
\left\langle\smash{#1}{\vphantom1}^{-}\right|{#2}%
\left|\smash{#3}{\vphantom1}^{-}\right\rangle}
\def\sandpp#1.#2.#3{%
\left\langle\smash{#1}{\vphantom1}^{+}\right|{#2}%
\left|\smash{#3}{\vphantom1}^{+}\right\rangle}
\def\sandpm#1.#2.#3{%
\left\langle\smash{#1}{\vphantom1}^{+}\right|{#2}%
\left|\smash{#3}{\vphantom1}^{-}\right\rangle}
\def\sandmp#1.#2.#3{%
\left\langle\smash{#1}{\vphantom1}^{-}\right|{#2}%
\left|\smash{#3}{\vphantom1}^{+}\right\rangle}

\def\trFive{{\rm tr}_5}

\def\NC{{N_c}}
\def\NF{{N_f}}
\def\CA{\mathcal{A}}

\def\Caravel{{\sc Caravel}}

\newcommand{\ii}{\,\mathrm{i}\,}
\newcommand{\gluon}{\mathrm{g}}
\newcommand{\quark}{\mathrm{q}}
\newcommand{\Quark}{\mathrm{Q}}
\newcommand{\sv}{\textrm{v}}
\newcommand{\pV}{p_{\sv}}
\newcommand{\pv}{\pV}

\begin{document}

\title{Leading-Color Two-Loop Amplitudes for Four Partons and a $W$ Boson in QCD}

\preprint{{
\footnotesize 
\begin{tabular}{l} 
  CERN-TH-2021-156\\ FR-PHENO-2021-12 \\ MPP-2021-181 
\end{tabular}
}}

\author[a,b,c]{S.~Abreu,}
\author[d]{F.~Febres Cordero,}
\author[e]{H.~Ita,}
\author[e]{M.~Klinkert,}
\author[a]{B.~Page,}
\author[f]{and V.~Sotnikov}

\affiliation[a]{Theoretical Physics Department, CERN, 1211 Geneva, Switzerland}
\affiliation[b]{Higgs Centre for Theoretical Physics, School of Physics and Astronomy,\\
The University of Edinburgh, Edinburgh EH9 3FD, Scotland, UK}
\affiliation[c]{Mani L. Bhaumik Institute for Theoretical Physics, Department of Physics and Astronomy,\\
UCLA, Los Angeles, CA 90095, USA}
\affiliation[d]{Physics Department, Florida State University, 77 Chieftan Way,
Tallahassee, FL 32306, USA}
\affiliation[e]{Physikalisches Institut, Albert-Ludwigs-Universit\"at Freiburg, \\
D-79104 Freiburg, Germany}
\affiliation[f]{Max Planck Institute for Physics (Werner Heisenberg Institute),
D--80805 Munich, Germany}

\abstract{
We present the leading-color two-loop QCD corrections for the scattering of four partons and a $W$ boson,
including its leptonic decay.
The amplitudes are assembled from the planar two-loop helicity amplitudes
for four partons and a vector boson decaying to a lepton pair, which are also used to determine the planar
two-loop amplitudes for four partons and a $Z/\gamma^*$ boson with a leptonic decay.
The analytic expressions are obtained by setting up a dedicated Ansatz 
and constraining the free parameters from numerical samples
obtained within the framework of numerical unitarity.
The large linear systems that must be solved to determine
the analytic expressions are constructed to be in Vandermonde form.
Such systems can be very efficiently solved, bypassing the bottleneck of Gaussian elimination.
Our results are expressed in a basis of one-mass pentagon functions,
which opens the possibility of their efficient numerical evaluation.
}

\maketitle


\section{Introduction}

The production of a $W$ boson in association with jets is a key process at hadron
colliders. It can be used for precise Standard-Model (SM) measurements and
allows to constrain models beyond the SM. The
$W+2$-jet signature is of particular interest, as it will play a distinguished role for future
precision QCD analyses. By now, many observables for this process have already
been measured at the LHC to 10\% relative
uncertainty~\cite{ATLAS:2014fjg,CMS:2017gbl}, and substantial improvements are
expected during Run 3 and the high-luminosity phase of the LHC. 
Theoretically this process stands out, because of the expected fast convergence 
in perturbation theory. Indeed,
at this jet multiplicity all production channels are already contributing at
Born level, with mild next-to-leading-order (NLO) QCD corrections over phase
space and a small sensitivity to higher-order corrections 
\cite{Campbell:2003hd,Berger:2009ep,Anger:2017nkq,Kallweit:2015dum,Azzurri:2020lbf}.  The calculation of the corresponding
next-to-next-to-leading-order (NNLO) QCD corrections is highly desirable,
as it will allow to assess the quality of higher-order perturbative predictions in
QCD~\cite{DiGiustino:2020fbk,Bonvini:2020xeo,Duhr:2021mfd}.

In this article we focus on the calculation of the two-loop scattering
amplitudes necessary to obtain the leading-color NNLO QCD corrections for the
production of a $W$ boson with two jets at hadron colliders. More precisely, we
present the analytic form of the gauge-invariant planar contributions to the
amplitudes for the scattering of four partons and a $W$ boson which decays into
leptons. These amplitudes are the leading-color contributions corresponding to
the formal large $N_c$ limit keeping the ratio $N_f/N_c$ fixed.  We retain at
the amplitude level the leptonic decay products of the vector boson, and
consider all possible helicity configurations of the external particles. We
focus on a representative physical kinematical channel for every amplitude that
we compute.  The presented amplitudes also give leading-color contributions for
the scattering of four partons and a $Z/\gamma^*$ boson which decays into
leptons, however, omitting the gauge invariant contributions originating from
$Z/\gamma^*$ coupling directly to a closed quark loop.

While the one-loop helicity amplitudes for  $e^+e^-\rightarrow 4$ partons
have been known for more than two decades~\cite{Bern:1997sc, Bern:1996ka,
Glover:1996eh, Campbell:1997tv}, no analytic expression for corresponding two-loop
scattering amplitudes has been available until now.  Nevertheless there has
been much related progress, first with a benchmark numerical evaluation of the
amplitudes under consideration~\cite{Hartanto:2019uvl} and second with the
computation of the two-loop squared amplitudes for the production of an on-shell
$W$ boson in association with a $b\overline{b}$ pair~\cite{Badger:2021nhg} 
(see also related work on the two-loop helicity amplitudes for the production of
a Higgs boson in association with a $b\overline{b}$ pair at the
LHC~\cite{Badger:2021ega}).

Progress in the field of two-loop scattering amplitudes in recent years has been
substantial. This progress has been made on two fundamental aspects of their calculation. First, there have
been important practical advances in understanding how to reduce scattering
amplitudes to master integrals through so-called integration-by-pars (IBP)
reduction~\cite{Chetyrkin:1981qh, Laporta:2000dsw}. This has come both in the
form of a better understanding of how to simplify IBP relations with 
unitarity-compatible 
techniques~\cite{Gluza:2010ws, Larsen:2015ped, Ita:2015tya}, as well as the
suggestion to perform the reduction numerically over finite fields, and
reconstruct the results from these evaluations~\cite{vonManteuffel:2014ixa,
Peraro:2016wsq}. By now a number of public implementations of reconstruction
methodologies exist~\cite{Klappert:2019emp, Klappert:2020aqs, Peraro:2019svx}.
In this context, an important advance has been the development of the multi-loop
numerical unitarity
method~\cite{Ita:2015tya,Abreu:2017xsl,Abreu:2017hqn,Abreu:2018zmy}, where
loop amplitudes are numerically computed by exploiting their analytic
properties.
On the second front, there has been a great deal of progress in understanding
and computing the required master integrals, specifically in the framework of
differential equations~\cite{Kotikov:1990kg, Remiddi:1997ny} in their canonical
form~\cite{Henn:2013pwa}. This powerful framework, when combined with Ansatz
techniques~\cite{Abreu:2018rcw}, has led to the computation of not only
the challenging planar two-loop five-point one-mass master
integrals~\cite{Papadopoulos:2015jft, Abreu:2020jxa, Canko:2020ylt} relevant for
the amplitudes discussed in this paper, but also the non-planar hexabox
integrals~\cite{Abreu:2021smk, Papadopoulos:2019iam}, which contribute for example to
subleading color contributions. 
Bases of multi-valued transcendental functions contributing to five-point one-mass scattering have been constructed in refs.~\cite{Badger:2021nhg,Chicherin:2021dyp}.
These are an important ingredient enabling the application of modern techniques for the calculation of scattering amplitudes.
Furthermore, there now exist multiple
methodologies for the numerical evaluation of the integrals in the form of
either generalized series expansions~\cite{Moriello:2019yhu, Hidding:2020ytt,Abreu:2020jxa,Badger:2021nhg},
or iterated integral methods in the form of so-called
\textit{pentagon functions}~\cite{Gehrmann:2018yef, Chicherin:2020oor, Chicherin:2021dyp}.

In this work, we make use of these recent advances, employing and extending them
to analytically compute for the first time the scattering amplitudes relevant for
the leading-color production of a $W$ boson decaying to leptons in association with two jets at the
LHC. In order to achieve this, we follow an Ansatz strategy, reconstructing the
analytic form of the amplitudes from numerical samples over finite fields. The
numerical reductions to master integrals are obtained with the numerical
unitarity approach~\cite{Ita:2015tya,Abreu:2017xsl,Abreu:2017hqn,Abreu:2018zmy}
using the implementations within the \Caravel{}
framework~\cite{Abreu:2020xvt}.
These numerical reductions, when combined with expressions
for the master integrals in terms of a basis of pentagon
functions~\cite{Chicherin:2021dyp}, allow us to subtract the known
infrared~\cite{Catani:1998bh, Becher:2009cu,Gardi:2009qi}
and ultraviolet pole contributions and perform a reconstruction of the
so-called \textit{finite remainder}. 
In order to efficiently perform the reconstruction, we describe a 
modern functional reconstruction algorithm. First we discuss how
to map the six-point kinematics of the amplitudes including the
$W$ boson decay to a lepton pair to the
(six-scale) five-point one-mass kinematics which underlie the QCD corrections to 
the process. Then we discuss how we
combine a univariate partial fraction approach~\cite{Badger:2021nhg} with an
approach built on Vandermonde-based sampling~\cite{Klappert:2019emp}, resulting
in a modern technique whose complexity is dominated by the sampling in the first finite field.

The paper is organized as follows. First, in section~\ref{sec:Notation} we establish the notation
to describe the amplitudes under consideration. Next, in section~\ref{sec:NumericalCalculation} we review the
numerical aspects of the calculation framework and in section~\ref{sec:reconstruction} we discuss
the modern functional reconstruction approach employed to determine the analytic expressions. We then
discuss implementation details and describe the analytic results in
section~\ref{sec:Results}.
The analytic expressions are provided in an accompanying set of ancillary
files \cite{W4partonsite}.
Finally, in section~\ref{sec:Conclusions} we summarize the
results of this paper. A series of appendices provide extra details of our
calculation and results.

trivial

\section{Notation and Conventions}
\label{sec:Notation}

The main results of this paper are the leading-color two-loop QCD helicity
amplitudes for the scattering of four partons and a $W$ boson,
where the $W$ boson decays into a lepton pair.
In the Standard Model, the $W$ boson couples with both vector and axial-vector couplings. 
However, it can be shown that if the $W$ boson always couples to a quark line connected to 
external states, as is the case in the amplitudes we consider, 
the vector and axial-vector contributions are equal up to an overall sign.
We discuss this relation in more detail in \cref{sec:gamma5}. For now
we simply note that this observation simplifies our computational setup. Indeed, it means that in order to 
assemble the amplitudes for four partons and a $W$ boson we can instead consider the
amplitudes for four partons and a vector boson $V$ which only interacts through a 
vector (and not axial-vector) coupling. 

\begin{figure}[]
  \centering
  \begin{subfigure}{0.4\textwidth}
  	\centering
  	\includegraphics[scale=0.5]{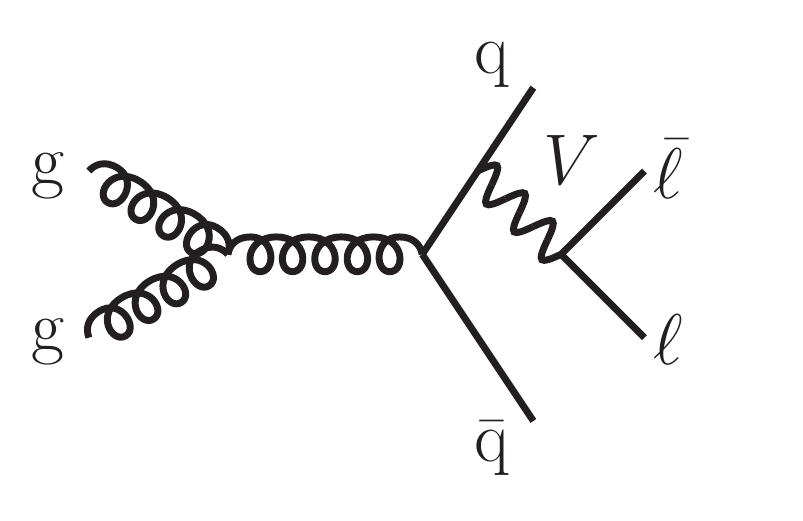}
  	\caption{Single quark line}
  \end{subfigure}
  \begin{subfigure}{0.4\textwidth}
  	\centering
  	\includegraphics[scale=0.5]{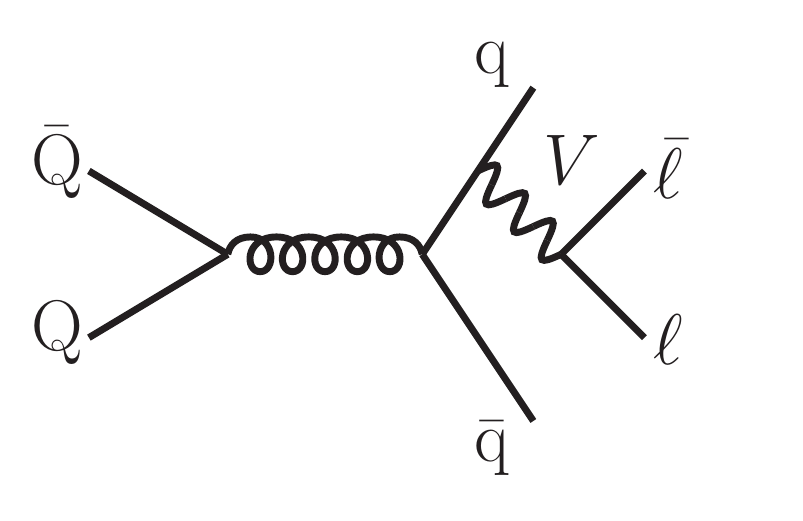}
  	\caption{Two quark lines}
  \end{subfigure}
\caption{Schematic representation of the two partonic processes
in eq.~\eqref{eq:partProcesses}}
\label{fig_schematic}
\end{figure}

Let us then consider the gauge-invariant planar contributions to the two-loop  QCD helicity
amplitudes for four partons and a vector boson $V$.
The vector boson only interacts through a vector current,
and decays into a lepton pair.
There are two independent partonic processes to consider,
one with a single quark line and one with two quark lines,
and we assume the vector boson only couples to one of the quark flavors, which we
denote as $\quark$.
In fig.~\ref{fig_schematic} we schematically depict the two processes at tree level.
The respective helicity amplitudes will be denoted as
\begin{align}\begin{split}\label{eq:partProcesses}
	&\mathcal{M}_\gluon\left(\bar \quark^{h_1}_{p_1},\gluon^{h_2}_{p_2},\gluon^{h_3}_{p_3},\quark^{h_4}_{p_4};
	\bar\ell^{h_5}_{p_5},{\ell}^{h_6}_{p_6}\right)\,,\\
	&\mathcal{M}_\Quark\left(\bar\quark^{h_1}_{p_1},\Quark^{h_2}_{p_2},\bar{\Quark}^{h_3}_{p_3},\quark^{h_4}_{p_4};
	\bar{\ell}^{h_5}_{p_5},{\ell}^{h_6}_{p_6}\right)\,,
\end{split}\end{align}
where $\gluon$ denotes a gluon, $\quark$ and $\Quark$ denote massless quarks which we assume
have different flavors, and we have singled out the pair of leptons $\ell$ and 
$\bar{\ell}$ which are produced in the decay of the vector boson $V$.
For each particle, we include a label $h_i$ to denote the helicity state and a 
label $p_i$ to denote the momentum.
In this paper we will focus on the particular physical partonic channels where $p_1$ and $p_2$ are incoming. 
The complete set of partonic channels required for phenomenological applications can be obtained by a straightforward application of permutations of particles' momenta, charge and parity conjugation (see discussion at the end of \cref{sec:results-validation}).

From the amplitudes in \cref{eq:partProcesses} we can then assemble the amplitudes
for Standard Model bosons.
In particular, since the $W$ boson cannot couple directly to a closed fermion loop,
the gauge-invariant planar contributions we consider in \cref{eq:partProcesses} capture
the full leading-color contributions for the amplitudes for four partons and a $W$ boson
decaying to leptons. 
Given that in this case the helicities of the emitting quark line and the leptons are fixed, we obtain, 
in all outgoing notation and closely following the notation of ref.~\cite{Bern:1997sc},
\begin{align}\begin{split}\label{eq:ampW}
\mathcal{M}^W\left( \bar u_{p_1}^R, g_{p_2}^{h_2}, g_{p_3}^{h_3}, d_{p_4}^L ,\bar e_{p_5}^R , \nu_{p_6}^L\right)
&= v^2\,{\cal P}_W(s_{56})\, \mathcal{M}_\gluon \left(\bar{q}^{+}_{p_1},g_{p_2}^{h_2},g_{p_3}^{h_3},{q}^{-}_{p_4};
  \bar{\ell}^{+}_{p_5},{\ell}^{-}_{p_6}\right)\,,\\
\mathcal{M}^W\left( \bar u_{p_1}^R, c_{p_2}^{h}, \bar{c}_{p_3}^{-h}, d_{p_4}^L ,\bar e_{p_5}^R , \nu_{p_6}^L\right)
&= v^2\,{\cal P}_W(s_{56})\, \mathcal{M}_\Quark \left(\bar{q}^{+}_{p_1},Q_{p_2}^{h},\bar{Q}_{p_3}^{-h},{q}^{-}_{p_4};
  \bar{\ell}^{+}_{p_5},{\ell}^{-}_{p_6}\right)\,,
\end{split}\end{align}
where $u$, $d$ and $c$ denote distinct quark flavours, and
\begin{equation}
{\cal P}_W(s)= \frac{s}{s-M_W^2+\ii\Gamma_W M_W} \,, \quad  
v^2=\frac{e^2}{2 \sin^{2}{\theta_w}}\,.
\end{equation}
$M_W$ is the mass of the $W$ boson, $\Gamma_M$ its decay width, 
$\theta_w$ is the Weinberg angle, and we assume a diagonal CKM matrix. 
For simplicity we only quote the result for quark lines from distinct families. 
The other combinations are obtained by suitable linear combinations.
We will see in \cref{sec:Zamp} how the amplitudes in \cref{eq:partProcesses}
can also be used to determine the planar gauge-invariant contributions for the
amplitudes for four partons plus a $Z/\gamma^*$ decaying to leptons.

Having identified the amplitudes in \cref{eq:partProcesses} as the central building
blocks, in the remaining of this paper we focus on their evaluation.
The calculation of the loop amplitudes is performed in the 't Hooft-Veltman (HV)
scheme of dimensional regularization  with $D=4-2\epsilon$ space-time dimensions.
The dependence of all quantities on the dimensional regulator $\epsilon$ will 
in general be kept implicit.
In dimensional regularization one must also take care with the precise
definition of the helicity states of the external quarks, and we refer the reader
to the detailed discussion in ref.~\cite{Abreu:2018jgq} where all our conventions
are described.

\begin{figure}[]
\centering
  \begin{subfigure}{0.31\textwidth}
  	\centering
  	\includegraphics[scale=0.5]{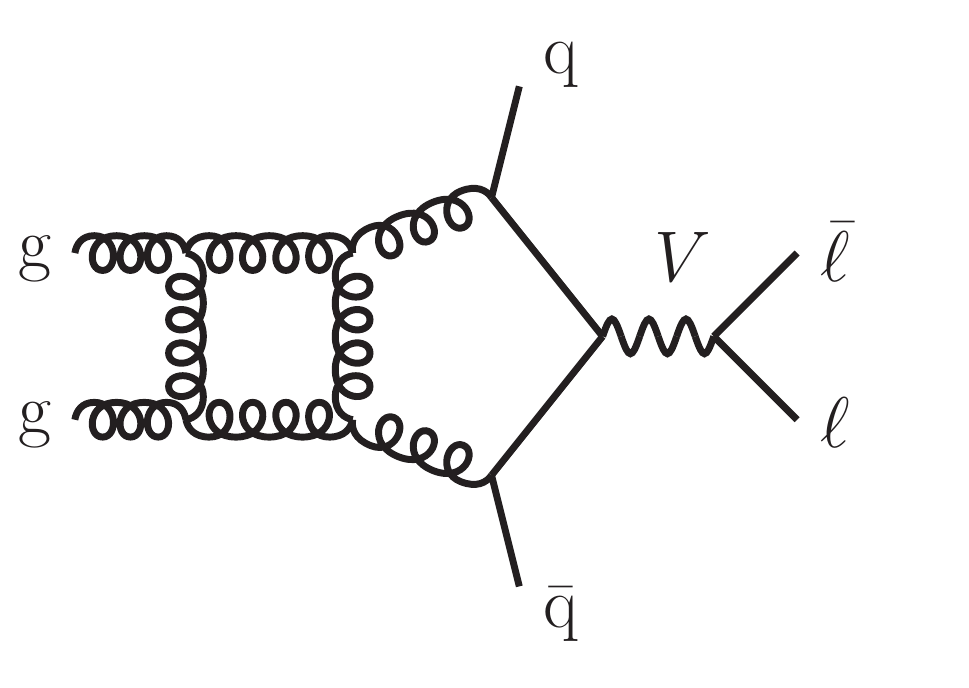}
  \end{subfigure}
  \begin{subfigure}{0.31\textwidth}
  	\centering
  	\includegraphics[scale=0.5]{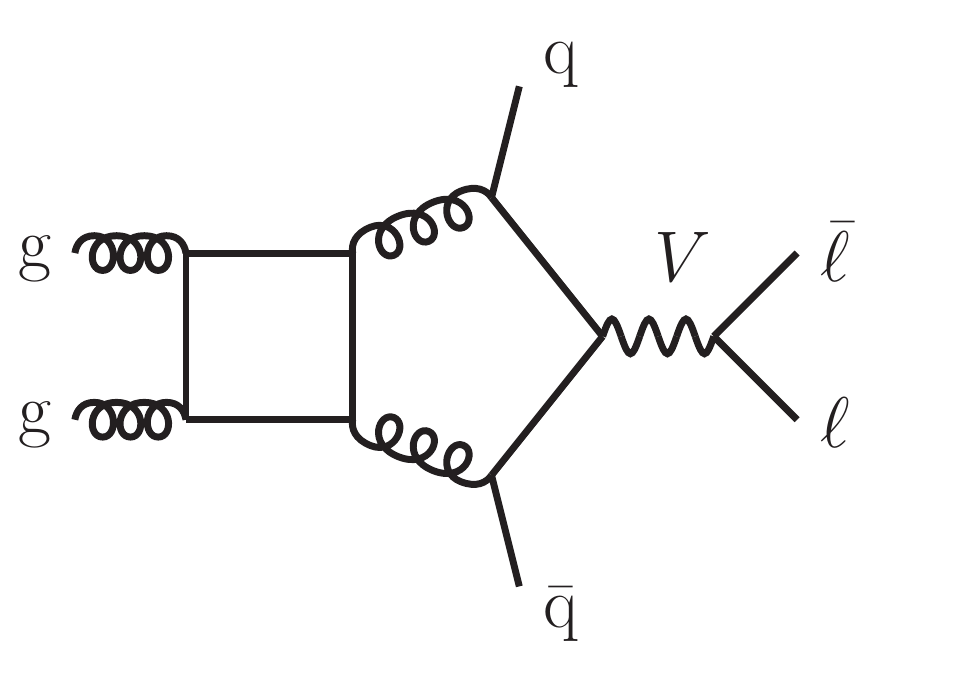}
  \end{subfigure}
   \begin{subfigure}{0.31\textwidth}
  	\centering
  	\includegraphics[scale=0.5]{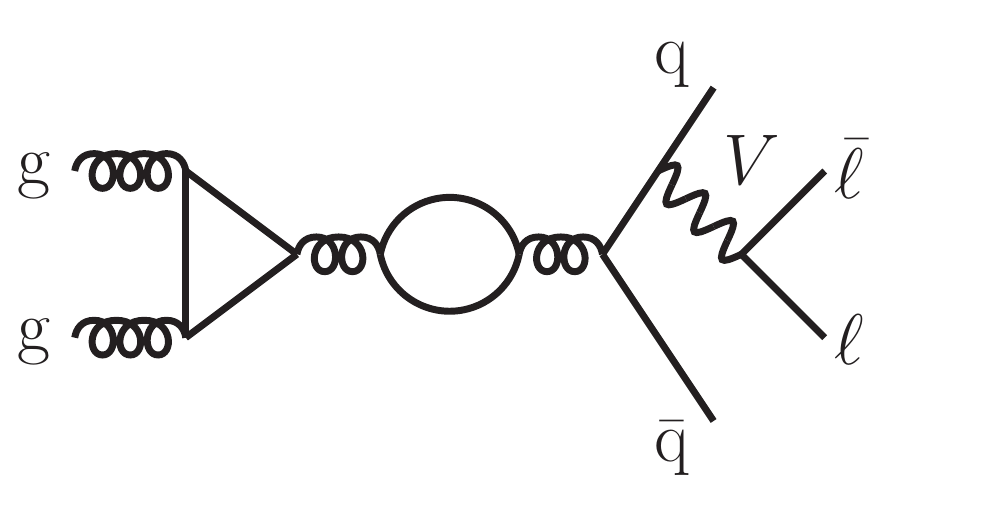}
  \end{subfigure}
\caption{Sample diagrams for the different contributions to 
\cref{eq:nfDecomp} for $\kappa=\gluon$.}
\label{fig_parents2q2g}
\end{figure}

\begin{figure}[]
\centering
  \begin{subfigure}{0.31\textwidth}
  	\centering
  	\includegraphics[scale=0.5]{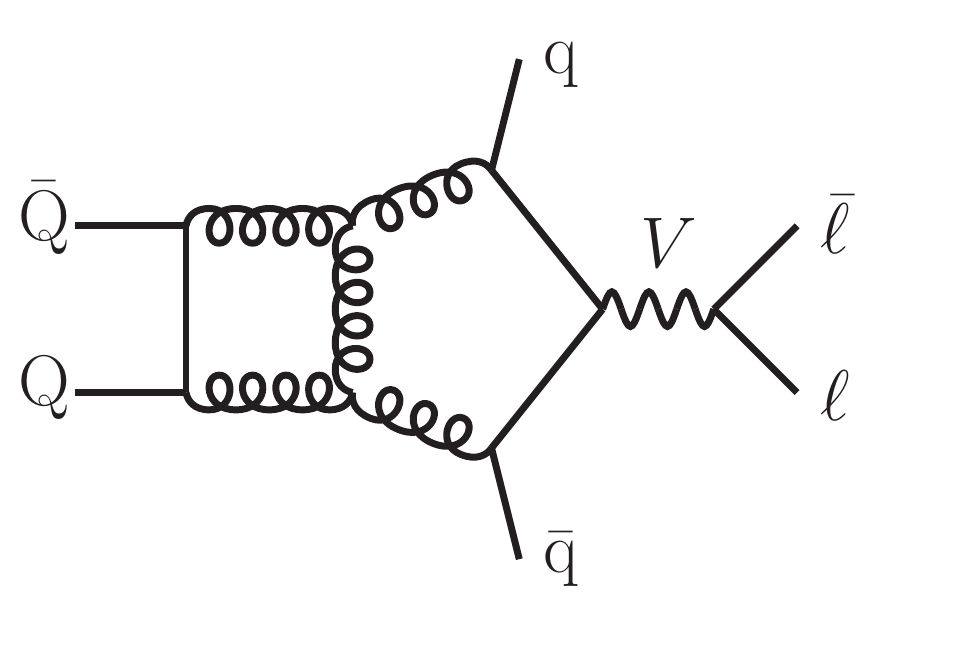}
  \end{subfigure}
  \begin{subfigure}{0.31\textwidth}
  	\centering
  	\includegraphics[scale=0.5]{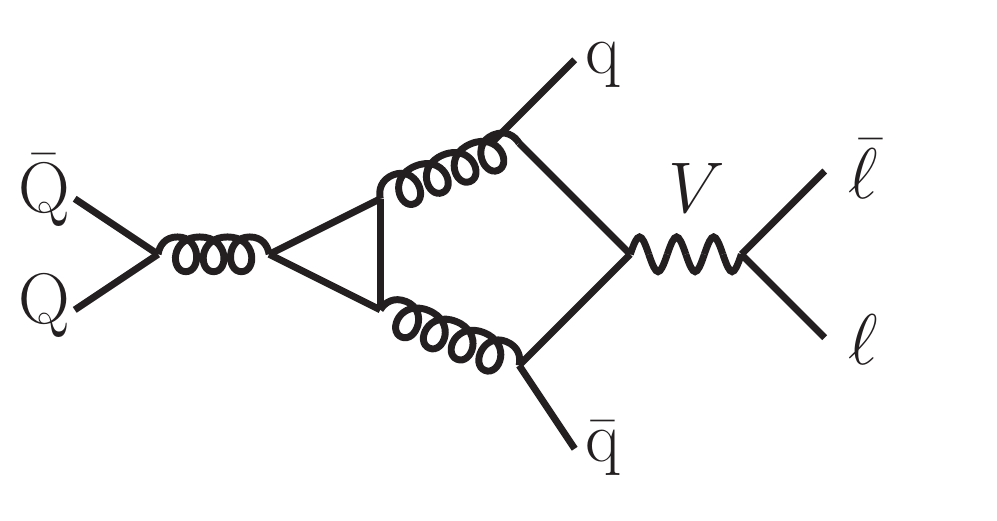}
  \end{subfigure}
   \begin{subfigure}{0.31\textwidth}
  	\centering
  	\includegraphics[scale=0.5]{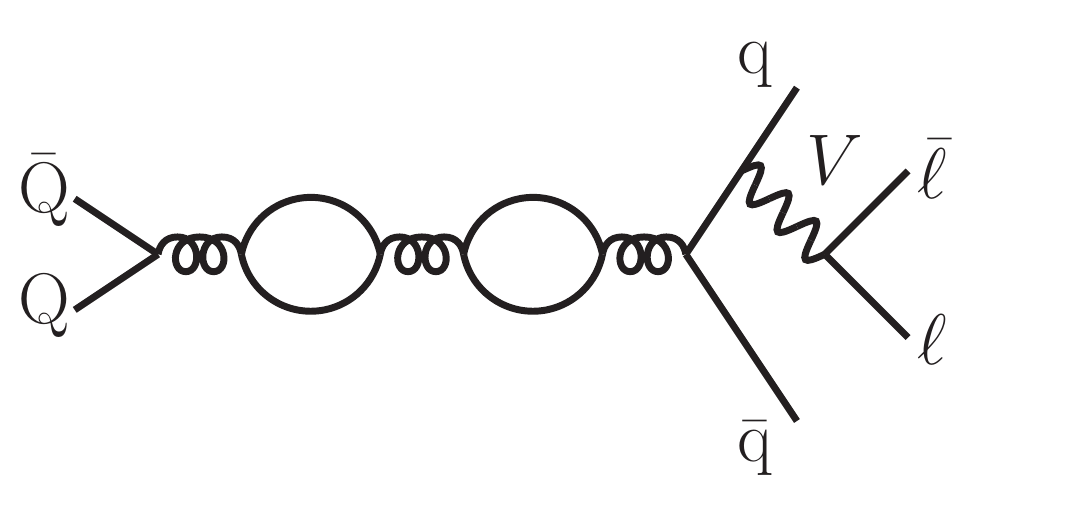}
  \end{subfigure}
\caption{Sample diagrams for the different contributions to 
\cref{eq:nfDecomp} for $\kappa=\Quark$.}
\label{fig_parents4q}
\end{figure}

The $\mathcal{M}_\gluon$ and $\mathcal{M}_\Quark$ amplitudes
can be expanded in powers of the bare coupling $\alpha_s^0=(g_s^0)^2/(4\pi)$,
\begin{equation}\label{eqn:loopSummedAmpl}
	\mathcal{M}_\kappa=(g_s^0)^2 \left(\mathcal{M}_\kappa^{(0)}+
	\frac{\alpha_s^0}{2\pi}\mathcal{M}_\kappa^{(1)}+
	\left(\frac{\alpha_s^0}{2\pi}\right)^2\mathcal{M}_\kappa^{(2)}+
	\mathcal{O}\left(\left(\alpha_s^0\right)^3\right)
	\right)\,,\qquad\textrm{for }\kappa=\gluon,\,\Quark\,.
\end{equation}
The tree level $\mathcal{M}_\kappa^{(0)}$ and one-loop $\mathcal{M}_\kappa^{(1)}$
contributions are known \cite{ Bern:1996ka,Bern:1997sc}, and in this paper we compute for the 
first time the two-loop
corrections $\mathcal{M}_\kappa^{(2)}$.
We also recompute the one-loop corrections~\cite{Bern:1997sc, Bern:1996ka}, and 
present results up to order $\epsilon^2$ in the dimensional regulator.
We consider the combinations of color factors and couplings
which receive contributions only from planar diagrams. 
In this framework, the color decomposition of the amplitudes is independent
of the loop order,
\begin{align}\begin{split}
	\mathcal{M}^{(k)}_\gluon&=\left(\frac{S_\epsilon\,\NC}{2}\right)^k\sum_{\sigma\in S_2}
        \left(T^{a_{\sigma(3)}}T^{a_{\sigma(2)}}\right)^{\,\,\bar{i}_1}_{i_4}\mathcal{A}_{\gluon}^{(k)}\,,\\
	\mathcal{M}^{(k)}_\Quark&=\left(\frac{S_\epsilon\,\NC}{2}\right)^k
        \delta_{i_2}^{\bar{i}_1}\delta_{i_4}^{\bar{i}_3}\mathcal{A}_{\Quark}^{(k)}\,,
\end{split}\end{align}
where  $S_\epsilon=(4\pi)^{\epsilon}e^{-\epsilon\gamma_E}\,$.
The partial amplitudes $\mathcal{A}_\kappa^{(k)}$ can be further decomposed into
powers of $\NF/\NC$, where $\NF$ denotes the number of massless quark flavors,
\begin{equation}\label{eq:nfDecomp}
	\mathcal{A}_\kappa^{(k)}=\sum_{j=0}^k
	\left(\frac{\NF}{\NC}\right)^j\mathcal{A}_\kappa^{(k)[j]}\,,\qquad\textrm{for }\kappa=\gluon,\,\Quark\,.
\end{equation}
We do not include any loop contributions from massive quark flavors.
In \cref{fig_parents2q2g,fig_parents4q} we present representative diagrams for the different
powers of $\NF$ for $\kappa=\gluon$ and $\kappa=\Quark$, respectively.

The renormalization of the amplitudes can be performed at the renormalization scale $\mu$
by replacing the bare coupling $\alpha^0_s$ by the renormalized coupling 
$\alpha_s(\mu)$ in the perturbative expansion of the 
bare amplitudes $\mathcal{M}_\kappa$. In the $\overline{\text{MS}}$ scheme, 
the bare coupling  is related to its renormalized counterpart by
\begin{equation}\label{eq:renormCoupling}
    \alpha_s^0\mu_0^{2\epsilon}S_{\epsilon}
  =\alpha_s(\mu)\mu^{2\epsilon}\left(
  1-\frac{\beta_0}{\epsilon}\frac{\alpha_s(\mu)}{2\pi}
  +\left(\frac{\beta_0^2}{\epsilon^2}-\frac{\beta_1}{2\epsilon}\right)
  \left(\frac{\alpha_s(\mu)}{2\pi}\right)^2+\mathcal{O}
  \left(\alpha_s^3(\mu)\right)\right),
\end{equation}
where  $\mu_0$ is the dimensional
regularization scale, which from now on we assume to be equal to $\mu$. We will
suppress the dependence on $\mu$ from all quantities. At leading color, and after setting 
$T_F=1/2$, $C_A=N_c$ and $C_F=(N_c^2-1)/(2N_c)$,
the coefficients of the QCD $\beta$-function are given by
\begin{equation}\label{eq:betai}
  \begin{aligned}
    \beta_0&=\frac{11 C_A - 4 T_F \NF }{6}=\frac{11N_c-2\NF}{6} \,,\quad\\
    \beta_1&=\frac{17C_A^2-6C_FT_F\NF-10C_AT_F\NF}{6}
    =\frac{1}{6}\left(17N_c^2-\frac{13}{2}N_c\NF\right)+\mathcal{O}(N_c^{-1})\,.
  \end{aligned}
\end{equation}
At each order in perturbation theory, the renormalized partial amplitudes are given by
\begin{align}\begin{split}
  \label{eq:twoLoopUnRenorm}
    &\mathcal{A}_{\kappa,R}^{(0)}=\mathcal{A}^{(0)}_\kappa, \\
  & \mathcal{A}_{\kappa,R}^{(1)}=\mathcal{A}_\kappa^{(1)}
  -\frac{2\beta_0}{\epsilon N_c}
  \mathcal{A}^{(0)}_\kappa\,,\\
  &\mathcal{A}_{\kappa,R}^{(2)}=
  \mathcal{A}^{(2)}_\kappa
  -\frac{4\beta_0}{\epsilon N_c}
  \mathcal{A}^{(1)}_\kappa
  +\left(\frac{4\beta_0^2}{\epsilon^2N_c^2}
  -\frac{2\beta_1}{\epsilon N_c^2}\right)
  \mathcal{A}^{(0)}_\kappa\,.
\end{split}\end{align}

As already stated above, our calculations are done in the HV scheme of dimensional
regularization and with the definition of helicity amplitudes from ref.~\cite{Abreu:2018jgq}.
To present our results in a form that is independent of these
choices (see e.g.~\cite{Broggio:2015dga}) and which is sufficient for physical applications
(see e.g.\ \cite{Weinzierl:2011uz}), we define
the finite remainders $\mathcal{R}_\kappa$. These are obtained by removing the
infrared singularities from the renormalized amplitudes, as they are
determined by the previous orders in perturbation theory and known universal 
factors~\cite{Catani:1998bh,Becher:2009cu,Gardi:2009qi}. The finite remainders
also admit an expansion in powers of the renormalized coupling,
\begin{equation}
  \mathcal{R}_\kappa = \mathcal{R}^{(0)}_\kappa + \frac{\alpha_s}{2\pi} \mathcal{R}^{(1)}_\kappa 
  + \left(\frac{\alpha_s}{2\pi}\right)^2 \mathcal{R}^{(2)}_\kappa + \mathcal{O}(\alpha_s^3)\,,
\end{equation}
with the $\mathcal{R}_\kappa^{(i)}$ defined as
\begin{align}\begin{split} \label{eq:remainder2l}
    \mathcal{R}^{(0)}_\kappa &= \mathcal{A}_{\kappa,R}^{(0)}, \\
    \mathcal{R}^{(1)}_\kappa &= \mathcal{A}_{\kappa,R}^{(1)}-\mathbf{I}_\kappa^{(1)}\mathcal{A}_{\kappa,R}^{(0)}
    ~+\mathcal{O}(\epsilon), \\ 
    \mathcal{R}^{(2)}_\kappa &= \mathcal{A}_{\kappa,R}^{(2)}-\mathbf{I}_\kappa^{(1)}\mathcal{A}_{\kappa,R}^{(1)}
    -\mathbf{I}_\kappa^{(2)}{\cal A}_{\kappa,R}^{(0)} ~+\mathcal{O}(\epsilon).
\end{split}\end{align}
For the amplitudes that we compute, the operators
$\mathbf{I}^{(1)}_{\kappa}$ and $\mathbf{I}^{(2)}_{\kappa}$ are 
diagonal in color space. 
For each $\kappa$, the operator $\mathbf{I}^{(1)}_{\kappa}$ is given by
\begin{align}
  {\bf I}^{(1)}_{\gluon}(\epsilon)=&
  -\frac{e^{\gamma_E\epsilon}}{\Gamma(1-\epsilon)}
  \left(
  \left(\frac{1}{\epsilon^2}+\frac{1}{\epsilon}\frac{\beta_0}{\NC}\right)
  \left( -s_{23}\right)^{-\epsilon}+
  \left(\frac{1}{\epsilon^2}+\frac{1}{\epsilon}\frac{\beta_0}{2\NC}+\frac{3}{4\epsilon}\right)
  \left(\left( -s_{12}\right)^{-\epsilon}+\left( -s_{34}\right)^{-\epsilon}\right)
  \right)\,,\nonumber\\
  {\bf I}^{(1)}_{\Quark}(\epsilon)=&
  -\frac{e^{\gamma_E\epsilon}}{\Gamma(1-\epsilon)}
  \left(\frac{1}{\epsilon^2}+\frac{3}{2\epsilon}\right)
  \left(\left( -s_{12}\right)^{-\epsilon}+\left( -s_{34}\right)^{-\epsilon}\right)\,,
\end{align}
where $s_{ij}=(p_i+p_j)^2$ and we take the Feynman prescription $s_{ij}
\rightarrow s_{ij} + i 0$ wherever relevant.
The operator~${\bf I}^{(2)}_{\kappa}$ is given by
\begin{align}\begin{split} \label{eqn:Iop}
    {\bf I}^{(2)}_{\kappa}(\epsilon)=&
  -\frac{1}{2}{\bf I}^{(1)}_{\kappa}(\epsilon)
  {\bf I}^{(1)}_{\kappa}(\epsilon)
  -\frac{2\beta_0}{\NC\epsilon}{\bf I}^{(1)}_{\kappa}(\epsilon) + 
  \frac{e^{-\gamma_E\epsilon}\Gamma(1-2\epsilon)}
  {\Gamma(1-\epsilon)}
  \left(\frac{2\beta_0}{\NC\epsilon}+K\right)
  {\bf I}^{(1)}_{\kappa}(2\epsilon) \\
  &+ \frac{e^{\gamma_E\epsilon}}{\epsilon\Gamma(1-\epsilon)}
  {\bf H}_{\kappa}\,,
\end{split}\end{align}
where 
\begin{equation}
K=\frac{67}{9}-\frac{\pi ^2}{3}-\frac{10}{9}\frac{\NF}{\NC}\,.
\end{equation}
The operator ${\bf H}_{\kappa}$
only depends on the number of external gluons and quarks:
\begin{equation}
  {\bf H}_{\gluon}=2H_\gluon+2H_\Quark\,,\qquad
  {\bf H}_{\Quark}=4H_\Quark\,,
\end{equation}
with
\begin{align}\begin{split}\label{eq:hexp}
  H_\gluon &= \left(\frac{\zeta_3}{2}+\frac{5}{12}+
  \frac{11\pi^2}{144}\right)
  -\left(\frac{\pi^2}{72}+\frac{89}{108}\right)\frac{N_f}{\NC}
  +\frac{5}{27}\left(\frac{N_f}{\NC}\right)^2\,,\\
  H_\Quark &=
  \left(\frac{7\zeta_3}{4}+\frac{409}{864}
  -\frac{11\pi^2}{96}\right)
  +\left(\frac{\pi^2}{48}-\frac{25}{216}\right)\frac{N_f}{\NC}\,.
\end{split}\end{align}

We note that the definition of remainders in \cref{eq:remainder2l} also 
subtracts contributions from the $\epsilon^0$ terms of
the Laurent expansion of $\mathcal{A}_{\kappa,R}^{(1)}$ and 
$\mathcal{A}_{\kappa,R}^{(2)}$.
The finite remainders can be expanded in powers of  $\NF/\NC$, 
in a similar way to
what was done for the partial amplitudes in \cref{eq:nfDecomp},
\begin{equation} \label{eq:remainder-nf-contrib}
\mathcal{R}^{(k)}_\kappa=
  \sum_{j=0}^{k}\left(\frac{N_f}{N_c}\right)^j
  \mathcal{R}^{(k)[j]}_\kappa\,.
\end{equation}

\subsection{Planar Amplitudes for Four Partons and a $Z/\gamma^*$ Boson}\label{sec:Zamp}

\begin{figure}[]
  \centering
  	\includegraphics[scale=0.5]{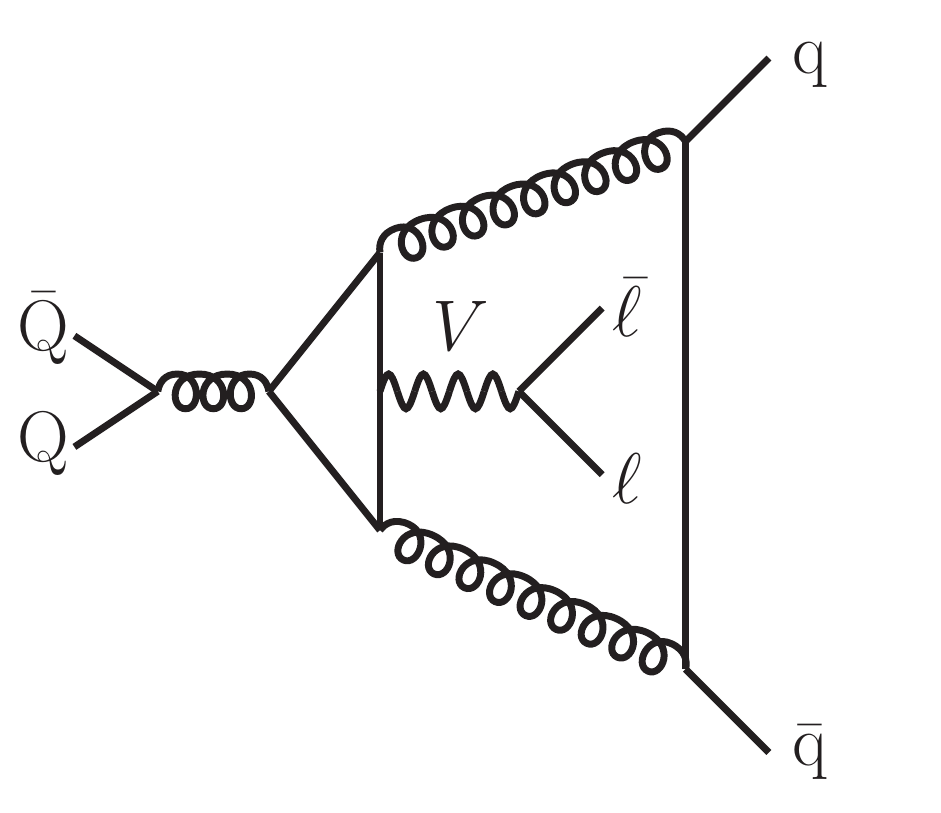}
  	\caption{A non-planar diagram that contributes in the leading-color 
	approximation to $Z$-boson production. This type of contribution has a 
	distinct coupling structure and is consistently dropped.}
\label{fig_np}
\end{figure}

In \cref{eq:ampW} we used the amplitudes in \cref{eq:partProcesses} as building blocks
to assemble the leading-color amplitudes for four partons and $W$ boson decaying into
a lepton pair. The same building blocks can be used to compute the planar contributions
for similar processes involving $Z/\gamma^*$ bosons decaying into a lepton pair.
As in \cref{eq:ampW}, this is achieved by combining the amplitudes of \cref{eq:partProcesses}
and dressing them with the corresponding couplings. The main difference with the $W$-boson
case is that in the $Z/\gamma^*$ case the leading-color contributions also contain non-planar
diagrams, as for example the one shown in \cref{fig_np}. Nevertheless, the planar
contributions that can be obtained from \cref{eq:partProcesses} form a well-defined 
gauge-invariant subset of the leading-color contributions, because the missing terms
can be associated to a distinct coupling structure.

Let us first consider the $Z/\gamma^*$ bosons decaying into a charged lepton pair. 
The corresponding amplitudes are given by
\begin{align}\begin{split}\label{eq:ampZ}
\mathcal{M}^{Z/\gamma^*}( \bar u_{p_1}^{-h}, & g_{p_2}^{h_2}, g_{p_3}^{h_3}, u_{p_4}^{h} ,\bar e^{-h'}_{p_5} , e^{h'}_{p_6})=\\
&= e^2 \big(- Q_q + v_e^{h'} v_q^{h} {\cal P}_Z(s_{56})\big)\mathcal{M}_\gluon \left(\bar{q}^{-h}_{p_1},g_{p_2}^{h_2},g_{p_3}^{h_3},{q}^{h}_{p_4};
	\bar{\ell}^{-h'}_{p_5},{\ell}^{h'}_{p_6}\right)\,,
\end{split}\end{align}
and
\begin{align}\begin{split}
\mathcal{M}^{Z/\gamma^*}( \bar u_{p_1}^{-h}, & d_{p_2}^{\tilde h}, \bar d_{p_3}^{-\tilde h}, u_{p_4}^h ,\bar e_{p_5}^{-h'} , e_{p_6}^{h'})=\\
&= e^2 \big(- Q_u + v_e^{h'} v_u^{h} {\cal P}_Z(s_{56})\big)\mathcal{M}_\Quark
\left(\bar{q}^{-h}_{p_1},Q_{p_2}^{\tilde h},\bar{Q}_{p_3}^{-\tilde h},{q}^{h}_{p_4};
	\bar{\ell}^{-h'}_{p_5},{\ell}^{h'}_{p_6}\right)\\
& \quad + e^2 \big( - Q_d + v_e^{h'} v_d^{\tilde h} {\cal P}_Z(s_{56})\big)\mathcal{M}_\Quark \left(\bar{q}^{-\tilde h}_{p_3},Q_{p_4}^{h},\bar{Q}_{p_1}^{-h},{q}^{\tilde h}_{p_2};
\bar{\ell}^{-h'}_{p_5},{\ell}^{h'}_{p_6}\right)\,,\label{eq:acont}
\end{split}\end{align}
where
\begin{align}\begin{split}
\mathcal{P}_Z(s) &=\frac{s}{s-M_Z^2+\ii\Gamma_Z M_Z} \,,\\
v_e^{-}&=v_e^{L} = \frac{ -1+2 \sin^2\theta_w}{\sin{2\theta_w}} \,, \quad 
v_e^{+}=v_e^{R} = \frac{ 2 \sin^2\theta_w}{\sin{2\theta_w}} \,, \\
v_{u,d}^{-}&=v_{u,d}^{L}=\frac{\pm 1-2 Q_{u,d} \sin^2{\theta_w}}{\sin{2 \theta_w}}\,,\quad 
v_{u,d}^{+}=v_{u,d}^{R}=\frac{-2 Q_{u,d} \sin^2{\theta_w}}{\sin{2 \theta_w}}\,.
\end{split}\end{align}
The $Q_{u,d}$ are respectively the charges of the up-/down-type quarks, $Q_u=2/3$ and $Q_d=-1/3$. 
The lepton pair that couples to the photon with charge $-1$ in units of $e$ is visible 
in the expression $-Q_{u/d}$. In $v_{u,d}^{h}$ 
the upper sign corresponds to the up-type coupling $v_{u}^{h}$ and
the lower one to the down-type coupling $v_{d}^{h}$. 

The $Z$ boson can also decay into a pair of neutrinos. In this case, we have 
\begin{align}\begin{split}\label{eq:ampZnus}
\mathcal{M}^Z( \bar u_{p_1}^{-h}, & g_{p_2}^{h_2}, g_{p_3}^{h_3}, u_{p_4}^{h} ,\bar \nu^{R}_{p_5} , \nu^{L}_{p_6})=\\
&=  e^2 v_\nu v_q^{h} {\cal P}_Z(s_{56}) \, \mathcal{M}_\gluon \left(\bar{q}^{-h}_{p_1},g_{p_2}^{h_2},g_{p_3}^{h_3},{q}^{h}_{p_4};
	\bar{\ell}^{+}_{p_5},{\ell}^{-}_{p_6}\right)\,,\\
\end{split}\end{align}
and
\begin{align}\begin{split}
\mathcal{M}^Z( \bar u_{p_1}^{-h}, & d_{p_2}^{\tilde h}, \bar d_{p_3}^{-\tilde h}, u_{p_4}^h ,\bar \nu_{p_5}^{R} , \nu_{p_6}^{L})=\\
&= e^2  v_\nu v_u^{h} {\cal P}_Z(s_{56}) \, \mathcal{M}_\Quark\left(\bar{q}^{-h}_{p_1},Q_{p_2}^{\tilde h},\bar{Q}_{p_3}^{-\tilde h},{q}^{h}_{p_4};
	\bar{\ell}^{+}_{p_5},{\ell}^{-}_{p_6}\right)\\
& \quad + e^2 v_\nu v_d^{\tilde h} {\cal P}_Z(s_{56}) \, \mathcal{M}_\Quark \left(\bar{q}^{-\tilde h}_{p_3},Q_{p_4}^{h},\bar{Q}_{p_1}^{-h},{q}^{\tilde h}_{p_2};
	\bar{\ell}^{+}_{p_5},{\ell}^{-}_{p_6}\right)\,,\label{eq:acont2}
\end{split}\end{align}
where
\begin{align}\begin{split}
v_\nu&= \frac{1}{\sin{2\theta_w}} \,.
\end{split}\end{align}


\section{Numerical Calculation of Amplitudes}
\label{sec:NumericalCalculation}

Our strategy for the computation of the loop amplitudes is to reconstruct
the analytic expressions from numerical samples. 
To simplify this process, we directly reconstruct the finite
remainders defined in \cref{eq:remainder2l}, that target the new contributions
at each loop order.
This is achieved by first numerically reducing the amplitude to a basis
of master integrals, which are themselves expanded in terms of a basis
of pentagon functions~\cite{Chicherin:2021dyp}. We can then subtract the known divergences and obtain a 
numerical decomposition of the remainders in terms of pentagon functions.

Before reviewing our strategy for the numerical computation of the amplitudes
and remainders, however, 
we start with a comment on the kinematics of the processes described by the amplitudes in \cref{eq:partProcesses}.
These processes depend on six massless momenta, $p_1$ through $p_6$. 
However, the amplitudes can be factorized into two contributions: 
the first describes the scattering of four partons and a (off-shell) vector boson, and the second
describes the decay of the vector boson into a pair of leptons. This is
apparent in the diagrams of \cref{fig_schematic,fig_parents2q2g,fig_parents4q}.
While $\mathcal{M}_\kappa$ depends on six massless momenta,
it is clear that the inherently two-loop parts of our calculation, the
amplitudes for one vector boson and four partons, only depend on 
five momenta $p_1$, $p_2$, $p_3$, $p_4$ and $\pV=p_5+p_6$, with $\pV^2\neq0$.
Even though this leads to a simplification compared to genuine six-point
kinematics, we note that it is nevertheless a significant
increase in complexity compared to previous calculations done
with the two-loop numerical unitarity approach. Indeed,
only five-point massless processes have so far been computed with it, and while the
algorithm itself is unaffected, some of the ingredients required at different
stages had to be updated. We will highlight these updates in
the following brief outline of our approach.

\subsection{Two-Loop Numerical Unitarity}\label{sec:2loopNumUni}

In order to perform the numerical reduction of the amplitudes to master
integrals we use the framework of two-loop numerical unitarity 
\cite{Ita:2015tya,Abreu:2017xsl,Abreu:2017hqn,Abreu:2018zmy,Abreu:2018jgq,Abreu:2020xvt}.
We begin by parametrizing the integrand
of the partial amplitudes ${\CA}^{(2)[j]}_\kappa(\ell_l)$ in terms of
master integrands, which integrate to master integrals, and surface terms, which
integrate to zero \cite{Ita:2015tya}.
Here $\ell_l$ collectively denotes the loop momenta of the problem.
Such a parametrization is most useful when organized in terms of propagator structures.
That is, in the numerical unitarity approach we write
\begin{equation}\label{eq:integrand}
  {\CA}^{(2)[j]}_\kappa(\ell_l)=\sum_{\Gamma\in\Delta}
  \sum_{i\in M_\Gamma\cup S_\Gamma}c_{\Gamma,i}
  \frac{m_{\Gamma,i}(\ell_l)}{\prod_{j\in P_\Gamma}\rho_j}\,,
\end{equation}
where $\Delta$ is the set of propagator structures $\Gamma$, and $P_\Gamma$
is the multiset of inverse propagators $\rho_j$ in $\Gamma$. $M_\Gamma$
and $S_\Gamma$ denote the sets of master integrands and surface terms $m_{\Gamma,i}(\ell_l)$ mentioned
previously, and $c_{\Gamma,i}$ are the corresponding coefficients. 
We construct a new set of surface terms from unitarity-compatible IBP identities \cite{Ita:2015tya,Gluza:2010ws,Schabinger:2011dz,Larsen:2015ped},
following an approach similar to the one employed in ref.~\cite{Abreu:2017hqn}, the details will be presented elsewhere \cite{maxThesis}.
We choose the set $M_\Gamma$ to correspond to pure master integrals from ref.~\cite{Abreu:2020jxa}, which
we cast into a form compatible with \cref{eq:integrand} employing \texttt{FiniteFlow}~\cite{Peraro:2019svx}.
In order to handle the significant increase of complexity of the sets $M_\Gamma$ and $S_\Gamma$,
compared to the previously considered case of five-point massless kinematics,
the implementation of the decomposition \eqref{eq:integrand} within \Caravel{} was improved.
This is the first time that the 
decomposition in \cref{eq:integrand} is obtained for processes that depend
on five-point one-mass kinematics. 

In order to determine the coefficients $c_{\Gamma,i}$
we rely on the factorization properties of the integrand. Specifically, we
consider ${\CA}_\kappa^{(2)[j]}(\ell_l)$ on loop-momentum configurations $\ell_l^\Gamma$
where the propagators are on-shell, that is $\rho_j(\ell_l^\Gamma) = 0$
iff $j\in P_\Gamma$. Taking such a limit, the leading contribution to
eq.~\eqref{eq:integrand} behaves as
\begin{equation}\label{eq:onshell}
  \sum_\text{states}\prod_{i\in T_\Gamma}{\CA}_i^{(0)}(\ell_l^\Gamma)=
  \sum_{\Gamma'\geq\Gamma,i\in M_{\Gamma'}\cup S_{\Gamma'}}
  \frac{c_{\Gamma',i}m_{\Gamma',i}(\ell_l^\Gamma)}
  {\prod_{j\in(P_{\Gamma'}\setminus P_\Gamma)}\rho_j(\ell_l^\Gamma)}\ .
\end{equation}
On the left-hand side of this equation, we denote by $T_\Gamma$ 
the set of tree amplitudes associated with the vertices in the diagram corresponding 
to $\Gamma$, and the sum is over the (scheme-dependent) physical states propagating through the internal
lines of $\Gamma$.  On the right-hand side, we sum over the propagator
structures which contribute to the limit, denoted $\Gamma'$, for which $P_\Gamma\subseteq P_{\Gamma'}$.

The coefficients $c_{\Gamma,i}$ can be determined numerically by sampling \cref{eq:onshell}
over a sufficient number of values of $\ell_l^\Gamma$. In order to work with
color-stripped products of tree amplitudes in \cref{eq:onshell} we make use of
the unitarity-based color decomposition approach of
refs.~\cite{Ochirov:2016ewn, Ochirov:2019mtf}. In order to handle the scheme dependence
inherent to the sum over states, which introduces a dependence on the dimensional
regulator on the left-hand side of \cref{eq:onshell}, we make use of the so-called
``decomposition by particle content'' approach ~\cite{Anger:2018ove,Abreu:2019odu,Sotnikov:2019onv},
based on dimensional reduction. 
We evaluate the tree amplitudes through Berends-Giele recursion \cite{Berends:1987me}.
In this way, we build a constraining system of equations for the $c_{\Gamma,i}$.
By performing all these calculations using finite-field arithmetic we are able to
determine the coefficients exactly with no loss of precision.
An important ingredient for this is a rational parametrization of
phase space, that can be obtained using the momentum-twistor parametrization of ref.~\cite{Hodges:2009hk}.
Having determined the $c_{\Gamma,i}$, we naturally arrive at the decomposition
of the amplitude in terms of master integrals,
\begin{equation}\label{eq:A}
  {\CA}^{(2)[j]}_\kappa =\sum_{\Gamma\in\Delta}
  \sum_{i\in M_\Gamma}c_{\Gamma,i}\mathcal{I}_{\Gamma,i}\,,
\end{equation}
where $\mathcal{I}_{\Gamma,i}$ is the master integral associated with the set of
propagators $\Gamma$ and numerator $m_{\Gamma,i}(\ell_l)$.
The set of master integrals relevant for this process are the planar two-loop five-point
one-mass integrals, for which we use the basis of ref.~\cite{Abreu:2020jxa}.

\subsection{Remainders and Pentagon Functions}

Our goal is not to stop at the decomposition of \cref{eq:A}, that is at the
decomposition of the amplitude in terms of master integrals. Rather, we
want to numerically evaluate the finite remainders of \cref{eq:remainder2l}.
Since the definition of the remainders involves different loop orders
and the operators $\mathbf{I}_\kappa^{(i)}$, it cannot be decomposed
into a set of master integrals. Instead, we consider
all sets of master integrals that appear in the definition of the finite remainders
and expand them order by order in the dimensional regulator $\epsilon$.
In such an expansion, the master integrals can be expressed in terms of multiple
polylogarithms (MPLs). MPLs are a class of special functions which generalize
the well-known logarithm and dilogarithm and, crucially, it is nowadays well understood
how to characterize all relations between
them~\cite{Goncharov:2010jf,Duhr:2011zq,Duhr:2012fh}. Therefore, it is possible
to construct a basis of the functions which arise in the master integrals order by order in $\epsilon$.
Functions in this basis are dubbed ``pentagon functions''. Such an analysis was made
in ref.~\cite{Chicherin:2021dyp}, alongside an efficient numerical implementation of the pentagon
functions, and we make use of this basis in our work.

Having evaluated the coefficients $c_{\Gamma,i}$ as described above, it
is a simple algebraic procedure to discard those associated to the surface terms
and insert expressions for the master integrals in terms of pentagon functions.
We thereby obtain a decomposition of the amplitude in terms of pentagon functions.
Denoting the relevant set of pentagon function monomials by $\{h_i\}_{i\in B}$, with $B$ the associated
set of labels, we can then write the amplitude as
\begin{equation}\label{eq:pent2l}
  {\CA}^{(2)[j]}_\kappa =\sum_{i\in B}\sum_{k=-4}^0\epsilon^k d_{k,i}h_i
  +\mathcal{O}(\epsilon)\,,
\end{equation}
where we make explicit that poles of order at most $\epsilon^{-4}$ can be found
in two-loop amplitudes (we suppress the indices $\kappa$ and $j$ on the right-hand side to avoid
overloading the notation).
As alluded to previously, by using the decomposition in \cref{eq:pent2l} 
and its one-loop equivalent, we are
able to write the one- and two-loop amplitudes in a common basis of functions.
Therefore, given \cref{eq:remainder2l,eq:remainder-nf-contrib}, the remainders can also
be decomposed in terms of pentagon functions. That is, we write
\begin{equation}\label{eq:pentFunc}
  \mathcal{R}^{(2)[j]}_\kappa =\sum_{i\in B}r_ih_i\,,
\end{equation}
where the $r_i$ are rational functions of the external kinematics (in which we
suppress for notational convenience the indices $\kappa$ and $j$).
In summary, we use the two-loop numerical unitarity approach to compute the
$r_i$ defined in \cref{eq:pentFunc} on a given phase-space point.
If desired, by using a rational parametrization of phase space, this 
calculation can be done over finite fields.


\section{Analytic Reconstruction Algorithm}
\label{sec:reconstruction}

As already discussed in the previous section, although we consider
six-point massless amplitudes, these are considerably simplified
by the way that the vector boson couples to the rest of the process.
Nevertheless, the underlying five-point one-mass kinematics makes this 
calculation considerably more complicated than the five-point massless 
amplitudes that have recently been the target of intense study.
In this section we discuss a reconstruction strategy which is able
to make use of the fact that the underlying kinematics of the corrections
we are computing are simpler than generic six-point massless kinematics,
and at the same time to handle the complex rational functions which arise.
Our strategy will combine a number of ideas proposed in the literature.
First, in ref.~\cite{Abreu:2018zmy} it was observed that the
denominators of rational functions can be systematically extracted from
knowledge of the coefficients on a univariate slice and the symbol 
alphabet.\footnote{Related methods have since been put forward \cite{Heller:2021qkz}.}
Second, in ref.~\cite{Badger:2021nhg} it was noticed that a
univariate partial fractioning of the coefficients simplifies their
analytic form in a way that is compatible with functional reconstruction
approaches.
Third, in ref.~\cite{Klappert:2019emp}, a simple method of efficiently reconstructing
polynomial functions was presented where a judicious choice of sample points
results in a structured linear system which can be efficiently solved.
Finally, it has by now become standard to exploit the simplifications taking
place when working at the level of the remainder, and so we take this as the
object of final interest in our discussion.

\subsection{Reduction to Five-Point One-Mass Kinematics}

The amplitudes under consideration in this work depend on the kinematics of
six massless particles. 
Since we will obtain analytic expressions with a functional reconstruction
strategy, for efficiency reasons it is crucial to work with the 
fewest possible variables.
In the following, we discuss how we reduce the six-point kinematics to an
underlying set of five-point one-mass kinematics. As the discussion holds for
all amplitudes $\mathcal{A}^{(i)[j]}_\kappa$ of \cref{eq:nfDecomp},
for the purpose of readability we shall
suppress indices which distinguish the various amplitudes.
We begin by noting that the amplitudes $\mathcal{A}$ factorize into a QCD
current $A^\mu$ and a tree-level leptonic current $J^\mu$, that is
\begin{equation}
  \mathcal{A} = A^\mu J_\mu, \quad \mathrm{where} \quad J_\mu = \overline{u}(p_6) \gamma_\mu v(p_5).
  \label{eq:CurrentFactorization}
\end{equation}
$\overline{u}$ and $v$ are Dirac spinors of definite helicity associated to 
the leptons $\bar\ell$ and $\ell$ in \cref{fig_schematic}, and
$A^\mu$ depends only on $p_5$ and $p_6$ through the combination $\pv = p_5 +
p_6$.
The QCD current $A^\mu$ is a Lorentz vector and we
need to specify a basis in which to express its components. 
As we work in the HV scheme, the contraction in equation
\eqref{eq:CurrentFactorization} is four dimensional 
(see e.g.~\cite{Gnendiger:2017pys}) and there are
exactly four components of the vector to be determined.
One of these four components can be trivially determined
by noting that both $A^\mu$ and $J^\mu$ satisfy the Ward identity
\begin{equation}
  \pv^\mu A_\mu = 0, \qquad \mathrm{and} \qquad \pv^\mu J_\mu = 0.
  \label{eq:WardIdentity}
\end{equation}
Therefore, only three components of $A^\mu$ remain to  be computed in order 
to specify it completely. 
In the following, we present a strategy to reduce the calculation of
these three components to that of on-shell 6-point scattering amplitudes in
constrained kinematics.

We begin by considering a set of kinematic configurations where one of the
leptons is collinear to one of the partons. This condition then fixes the
momenta of both of the leptons via on-shellness and momentum conservation. We
denote the configuration where $p_5$ is collinear to the momentum $p_i$ as
\begin{align}
p_5^{(i)} = \frac{\pv^2}{2p_i \cdot \pv} p_i,\ \ \ p_6^{(i)} = \pv- \frac{\pv^2}{2p_i\cdot \pv} p_i.
  \label{eq:LeptonicCollinearity}
\end{align}
Let us stress that due to the factorization in \cref{eq:CurrentFactorization},
the amplitude has
no singularity on such configurations, spurious or otherwise. In practice, we
will compute the amplitude in three such collinear configurations, specifically
where $i = 1, 2, 3$. 
This choice is arbitrary but we found it suitable for our reconstruction
strategy to succeed.
We denote the amplitude computed on these collinear
configurations as $\mathcal{A}^{\{i\}}$,
\begin{equation}
  \mathcal{A}^{\{i\}} = \mathcal{A}\left(p_1, p_2, p_3, p_4, p_5^{(i)}, p_6^{(i)}\right), \qquad i = 1, 2, 3.
\end{equation}
We dub the $\mathcal{A}^{\{i\}}$ form factors, as it is possible to recover the
QCD current $A^\mu$ from the $\mathcal{A}^{\{i\}}$. 
Indeed, this can be achieved by
introducing a judicious partition of
unity into \cref{eq:CurrentFactorization} as we will now make explicit. 
First, we introduce three reference directions defined as the leptonic
current $J^\mu$ evaluated in each collinear configuration, that is
\begin{equation}
   n_i^\mu = J^\mu\left(p_5^{(i)}, p_6^{(i)}\right), \qquad i = 1, 2, 3,
\end{equation}
and we furthermore define $n_4^\mu = \pv^\mu$, which will allow us
to directly use \cref{eq:WardIdentity} to determine one of the 
components of $A^\mu$. 
With these four reference directions,
we decompose the four-dimensional metric tensor $g^{\mu \nu}_{(4)}$ as 
\begin{equation}
    g^{\mu \nu}_{(4)} = \sum_{i,j=1}^4 G^{-1}_{ij} n_i^\mu n_j^\nu, \qquad \mathrm{where} \qquad G_{ij} = n_i \cdot n_j.
    \label{eq:MetricDecomposition}
\end{equation}
Due to the Ward identity in \cref{eq:WardIdentity}, $G_{i4} = G_{4j} =
0$ for $i, j = 1,2,3$.
Employing the partition of unity \eqref{eq:MetricDecomposition}, we can now
write the QCD current $A^\mu$ as
\begin{equation}
  A^\mu = \sum_{i,j=1}^3 G^{-1}_{ij} \mathcal{A}^{\{i\}} n_j^\mu,
\end{equation}
where we have used the fact that $\mathcal{A}^{\{i\}} = A \cdot n_{i}$, for $i =
1, 2, 3$ and \cref{eq:WardIdentity} to drop the $n_4$ contribution. It
is therefore sufficient to compute the form factors $\mathcal{A}^{\{i\}}$,
which are obtained as on-shell six-point amplitudes, in order to extract
the QCD current $A^\mu$.

There is a technical subtlety in applying a functional reconstruction
procedure to determine the $\mathcal{A}^{\{i\}}$: they are
not little group invariant, and hence depend on more variables than 
Mandelstam invariants. This is easily remedied by factoring
out some function with the same little group weights. Specifically, we make use
of a standard spinor weight defined in \Caravel{}, see Appendix A.3 of 
ref.~\cite{Abreu:2020xvt}. 
For the purpose of exposition, we shall suppress this
in the rest of the discussion and regard $\mathcal{A}^{\{i\}}$ as a rational
function of 
\begin{equation}
  \vec{s} = \{s_{\sv 1}, s_{12}, s_{23}, s_{34}, s_{4\sv}, \pv^2 \} \quad \mathrm{and} \quad \mathrm{tr}_5 = 4 \ii \varepsilon_{\mu_1\mu_2\mu_3\mu_4} p_1^{\mu_1} p_2^{\mu_2} p_3^{\mu_3} p_4^{\mu_4},
  \label{eq:FivePointOneMassVariables}
\end{equation}
where $s_{ij} = (p_i + p_j)^2$, and $\varepsilon_{\mu_1\mu_2\mu_3\mu_4}$ is the fully antisymmetric Levi-Civita symbol, $\varepsilon_{0123}=1$.

To close the discussion on the reduction from six-point massless to five-point
one-mass kinematics, we note that it is straightforward to define
an analogue of the form factors $\mathcal{A}^{\{i\}}$ for the finite
remainders. That is, we define
\begin{equation}
  \mathcal{R}^{\{i\}} = \mathcal{R}\left(p_1, p_2, p_3, p_4, p_5^{(i)}, p_6^{(i)}\right)\,, \qquad i = 1, 2, 3,
  \label{eq:RemainderFormFactors}
\end{equation}
implicitly considered to be normalized by the \Caravel{} 
spinor weight. The $\mathcal{R}^{\{i\}}$ are sufficient to reconstruct the full
finite remainders defined in \cref{eq:remainder2l}.
We will target our analytic reconstruction procedure on the form-factor
remainders $\mathcal{R}^{\{i\}}$, which
can be decomposed in terms of pentagon functions similarly 
to \cref{eq:pentFunc},
\begin{equation}\label{eq:pentFuncFF}
  \mathcal{R}^{\{i\}} =\sum_{i\in B}r_i(\vec{s}, \mathrm{tr}_5)h_i\,,
\end{equation}
where the $r_i(\vec{s}, \mathrm{tr}_5)$ are rational functions of 
their arguments.

\subsection{Common-Denominator Form}\label{sec:cdAnsatz}

Having reduced the problem of obtaining analytic expressions for the
amplitudes in \cref{eq:partProcesses} to a functional
reconstruction problem in the six variables of
\cref{eq:FivePointOneMassVariables}, we shall now discuss the first step
of our functional reconstruction strategy, which will allow us to gauge the
complexity of the problem we are trying to solve. 

We begin by reviewing some basic properties of the rational functions 
appearing in the finite remainders defined in \cref{eq:RemainderFormFactors}.
As noted earlier, the amplitudes are normalized such that the functions $r_i$
in \cref{eq:pentFuncFF} are rational functions of the Mandelstam
invariants and $\mathrm{tr}_5$. Furthermore, as $\mathrm{tr}_5^2$ is a
polynomial in the Mandelstam invariants, the dependence on it simplifies
and one can write
\begin{equation}\label{eq:parityDecomposition}
  r_i(\vec{s}, \mathrm{tr}_5) = r_{i}^+(\vec{s}) + \mathrm{tr}_5 \, 
  r_{i}^-(\vec{s})\,,
\end{equation}
where the $r_{i}^\pm$ are rational functions of the Mandelstam invariants. It is
clear that we can evaluate the $r_i^\pm$ by computing the $r_i$ on 
parity-conjugate phase-space points~\cite{Abreu:2019odu}.
As first suggested in ref.~\cite{Abreu:2018zmy},
when considered in common-denominator form the denominators of the
$r_{i}^\pm$  factorize into 
products of symbol letters raised to some power. We
take this form as an Ansatz and write
\begin{equation}
  r_{i}^\pm = \frac{n_{i}^{\pm}}{\prod_{j=1}^{37} W_j^{q_{ij}^\pm}},
  \label{eq:CommonDenominatorForm}
\end{equation}
where the $W_j$ are the 37 Galois-invariant letters of the planar
five-point one-mass symbol alphabet~\cite{Abreu:2020jxa}, and the exponents
$q_{ij}^{\pm}$ are (potentially negative) integers.
We note that this property holds if we normalize by the \Caravel{} 
spinor-weight factor, but not necessarily if we normalize by the corresponding
tree  amplitudes.

To determine the exponents $q_{ij}^{\pm}$ we use the method of
ref.~\cite{Abreu:2018zmy}. That is, we reconstruct the amplitude on a generic
univariate slice of kinematic space and match the factors on the univariate
slice against the symbol alphabet, performing all calculations in 
a finite field.
In order to keep the univariate reconstruction as simple as possible, we desire
a slice of phase space on which the Mandelstam behave linearly on 
the free parameter parametrizing the slice.
In particular, this guarantees that they
have the lowest possible non-trivial polynomial degree.
The construction of a set of momenta corresponding to such a univariate
slice is a subtle problem as it requires rationalizing $\mathrm{tr}_5$.
Such a parametrization
can be constructed on a case-by-case basis, making use of momentum-twistor
parametrizations of phase space.
Here we discuss an alternative and generic method of constructing a univariate slice of phase
space where Mandelstam variables are linear, based on generalized BCFW shifts.
Specifically, we propose to use a multi-line purely holomorphic
shift~\cite{Risager_2005}, as this naturally results in Mandelstam variables 
that are linear in the shift parameter. 
In order to be sufficiently generic, we shift all of the lines 
(see e.g.~ref.~\cite{Elvang:2008vz}) as  proposed in 
ref.~\cite{PageSAGEXLectures}. 
We begin with a generic momentum configuration
specified in terms of spinors
$\{\lambda_1,...,\lambda_6,\tilde{\lambda}_1,...,\tilde{\lambda}_6\}$. An
all-line holomorphic shift adjusts every $\lambda$ spinor in a way which is
proportional to a common reference spinor $\eta$, that is
\begin{align}
\lambda_i \rightarrow \lambda_i+t c_i \eta.
\end{align}
Here, the $c_i$ are proportionality factors designed to ensure that 
the shifted kinematics satisfy momentum conservation, which is ensured
by asking that they solve
\begin{align}
\sum_{i=1}^6 c_i\tilde{\lambda}_i=0\,.
  \label{eq:AllLineShiftConstraint}
\end{align}
This equation does not have a unique
solution but any solution is suitable.
Under such a shift, the holomorphic spinor products transform 
linearly in $t$
\begin{align}
  \langle ij \rangle \rightarrow \langle ij \rangle + t \, \big(c_i \langle \eta j \rangle + c_j \langle i \eta \rangle \big),
\end{align}
whereas the anti-holomorphic spinor products remain unchanged. 
It is therefore easy
to see that the Mandelstam invariants become linear in $t$ and the pseudo-scalar
$\mathrm{tr}_5$ is quadratic in $t$.
In practice, we pick the initial set of spinor variables and the $c_i$ 
randomly over the finite field, and hence the slice we generate in this way 
is generic.
We first generate shifted six-point kinematics, and then redefine
$p_5$ and $p_6$ so that $p_5$ satisfies the collinearity condition specified 
in~\cref{eq:LeptonicCollinearity}.

By reconstructing the coefficients $r_{i}^\pm$ on a univariate slice obtained
with this approach, we are able to determine the exponents $q_{ij}^\pm$
in \cref{eq:CommonDenominatorForm}. The degree of the polynomials $n_{i}^{\pm}$
can also be determined, and in principle we could then write an Ansatz for the $n_i^\pm$ as
a polynomial of that degree.
For the functions we are interested in, 
the degree of the numerator polynomial can be sufficiently
high that this Ansatz has tens of millions of free parameters.
The complexity involved in solving the linear systems required to constrain such an Ansatz
makes this an inviable approach to determining the analytic form of the $r_{i}^\pm$.

\subsection{Partial-Fraction Ansatz}\label{sec:partFracAns}

In order to address the problem identified above and
identify simpler polynomials to reconstruct from numerical data, 
we employ a univariate partial-fraction
decomposition as suggested in ref.~\cite{Badger:2021nhg}. 
In the following, we review this approach in order to discuss the details of 
our implementation.

We choose to partial fraction with respect to the Mandelstam
variable $s_{34}$. This choice is made arbitrarily and it would be interesting
to systematically investigate the effects of choosing other variables.
To facilitate the discussion, we introduce a vector of the remaining variables,
that is we define
\begin{equation}
  \vec{s}_{\mathrm{rem}} = \left\{s_{\sv 1}, s_{12}, s_{23}, 
  s_{4\sv}, \pv^2 \right\}.
\end{equation}

We begin by reviewing the systematics of univariate partial fractioning.
We start with the common-denominator representation of the rational function
\cref{eq:CommonDenominatorForm}, in which we recall the denominator is known
fully analytically,
and consider the situation where the
denominator contains two factors, $W_{j_1}^{Q_{j_1}}$ and $ W_{j_2}^{Q_{j_2}}$,
where both $W_j$ are non-constant functions of~$s_{34}$.
We are able to break the denominator into
two pieces using a solution to the equation
\begin{equation}
  1 = c_{j_1}(\vec{s}_{\mathrm{rem}}, s_{34}) W_{j_1}^{Q_{j_1}} +   c_{j_2}(\vec{s}_{\mathrm{rem}}, s_{34}) W_{j_2}^{Q_{j_2}},
  \label{eq:univariateSplitting}
\end{equation}
where the $c_j$ are polynomial in $s_{34}$ and rational in the
$\vec{s}_{\mathrm{rem}}$. 
The existence of such solutions to \cref{eq:univariateSplitting} is 
guaranteed by
Hilbert's Nullstellensatz as $W_{j_1}$ and $W_{j_2}$ are distinct irreducible
polynomials in $s_{34}$.
Equation \eqref{eq:univariateSplitting}
provides a decomposition of unity that we can use to break a term with
multiple $s_{34}$-dependent denominators into two terms, where each term 
has fewer such denominators.
Having started from  \cref{eq:CommonDenominatorForm},
we can repeatedly apply \cref{eq:univariateSplitting} in order to arrive at
a representation where each term has only a single denominator which depends on
$s_{34}$.
Once this has been achieved, to finally reach a canonical partial-fraction
decomposition, the second necessary step is to cancel numerators against the
denominators in $s_{34}$. This is easily achieved by polynomial division.
In general this procedure will generate spurious poles, which are polynomials 
in $\vec{s}_{\mathrm{rem}}$ and come in two classes. The first come from the
polynomial division step and
are the coefficients of the leading power of $s_{34}$ in the $W_j$.
They are trivial to generate. The
second come from the splitting step and are the denominators of the $c_i$.
To generate them, we consider all possible pairs of
$W_j$, construct the associated pairs of $(c_{j_1}, c_{j_2})$, and take their
denominators. We achieve this systematically by constructing the $(c_{j_1},
c_{j_2})$ pairs via Groebner basis methods, where $(c_{j_1}, c_{j_2})$ are
recognized as the co-factors arising in the computation of the Groebner basis of $W_{j_1}^{Q_{j_1}}$ and $W_{j_2}^{Q_{j_2}}$.
We finally collect all such spurious poles
together with all $s_{34}$ independent Galois-invariant letters
into the set $\overline{\mathcal{W}}=\{\overline{W}_1(\vec{s}_{\mathrm{rem}}), \ldots,
\overline{W}_{|\overline{\mathcal{W}}|}(\vec{s}_{\mathrm{rem}}) \}$.

We can now describe the details of our
partial-fraction decomposition of the $r_i^{\pm}$.
We parametrize our partial-fraction decomposition as
\begin{align}
    r_{i}^{\pm} \!=\! \sum_{j=1}^{n_i} \!P_{ij}^{\pm}(\vec{s}_{\mathrm{rem}} ) \, g_{ij}^{\pm}(\vec{s}_{\mathrm{rem}}, s_{34}), \quad \!\mathrm{with}\!
  \quad g_{ij}^{\pm}(\vec{s}_{\mathrm{rem}}, s_{34}) \!=\! \frac{Q^{\pm}_i(\vec{s}_{\mathrm{rem}}, s_{34}) s_{34}^{\alpha_{ij}^{\pm}}}{W_{k_{ij}}(\vec{s}_{\mathrm{rem}}, s_{34})^{\beta_{ij}^{\pm}} \prod_l \overline{W_l}(\vec{s}_{\mathrm{rem}})^{\gamma_{ij l}^{\pm}}}.
  \label{eq:PartialFractionDecomposition}
\end{align}
Here, the $P_{ij}^{\pm}$ are polynomial in $\vec{s}_{\mathrm{rem}}$, while
$g_{ij}^{\pm}$ contains all of the pole structure
discussed above, both physical and spurious. 
The functions $Q_i^{\pm}$ are products of the $W_j$ in 
\cref{eq:CommonDenominatorForm} which are excluded from
the partial-fraction analysis. The factors to be excluded are determined
experimentally by requiring that they lower the degree of the polynomials $P_{ij}^{\pm}$.
It would certainly be interesting to systematize this step of the procedure.
The index $l$ runs over all elements of the set $\overline{\mathcal{W}}$
defined above.
The exponents $\alpha_{ij}^{\pm}, \beta_{ij}^{\pm}, \gamma_{ijl}^{\pm}$ are all 
(potentially negative) integers.
Each term in the sum in \cref{eq:PartialFractionDecomposition} represents a 
term in the partial-fraction decomposition. 
Therefore, other than in the excluded piece $Q_i^{\pm}$, only a
single denominator factor $W_{k_{ij}}$ depends on $s_{34}$.

We view \cref{eq:PartialFractionDecomposition} as an Ansatz that should now be constrained
from numerical data. The $Q_i^{\pm}$, $W_{k_{ij}}$ and $\overline{W_l}$ are known fully
analytically, and there are two types of undetermined contributions: the polynomials $P_{ij}^{\pm}$, 
and the exponents $\alpha_{ij}^{\pm}$, $\beta_{ij}^{\pm}$ and $\gamma_{ijl}^{\pm}$.

We first discuss how to determine the exponents $\alpha_{ij}^{\pm}$, $\beta_{ij}^{\pm}$ 
and $\gamma_{ijl}^{\pm}$, as well as the total degrees of the polynomials $P_{ij}^{\pm}$.
This can be achieved in a very similar way to how the exponents in the common denominator form
of \cref{eq:CommonDenominatorForm} were determined. Instead of a univariate slice, however, 
we reconstruct the amplitude on a bivariate slice of phase space -- one
where $s_{34}$ varies freely and the other Mandelstam variables are linear in a
parameter. Specifically, we set
\begin{equation}\label{eq:bivariateSlice}
  s_{34}=s , \quad  s_{\sv 1} = a_1 + b_1 t, \quad \ldots, \quad 
  s_{4\sv} = a_4 + b_4 t, \quad \pv^2 = a_5 + b_5 t,
\end{equation}
where the $a_i$ and $b_i$ are fixed, randomly chosen elements of 
a finite field that define the slice.
At this stage we again face the problem that Mandelstam invariants do not 
rationally parametrize the phase space. 
To handle this difficulty, we begin with a parametrization of phase space which
rationalizes the Mandelstam variables and $\mathrm{tr}_5$, which we detail in
\cref{sec:ratParam}. This parametrization depends on the five Mandelstam
variables $\vec{s}_{\mathrm{rem}}$ and a final variable $x$.
The remaining Mandelstam invariant $s_{34}$ and $\mathrm{tr}_5$ depend
rationally on these variables. We note that the parametrization is tuned such
that the five $t$-dependent Mandelstam variables are explicitly independent. 
It is therefore trivial to choose $\vec{s}_{\mathrm{rem}}$ 
with a $t$ dependence as in~\cref{eq:bivariateSlice}.
To obtain the dependence on $s$, we apply the methodology of 
ref.~\cite{Abreu:2019odu}, where a change of variables is introduced in the
intermediate stages of the reconstruction procedure.
That is, we sample over values of $\{t, x\}$, and reconstruct the
rational functions through their dependence on 
$\{ s(t,x),t\}=\{s_{34}(t,x) , t\}$.
This approach also requires the ability to find parity conjugate phase-space
points, which we also detail in \cref{sec:ratParam}.
By reconstructing the amplitude on such a slice, and performing a univariate 
partial-fraction decomposition in $s$, we can evaluate the general Ansatz 
of~\cref{eq:PartialFractionDecomposition} on the bivariate slice.
One can then easily identify each term in the sum of
\cref{eq:PartialFractionDecomposition}, as they are uniquely identified by their
$s_{34}$ dependence. It is then elementary to extract the degrees of
$P_{ij}^\pm$ as well as the exponents of the factors of the $g_{ij}^{\pm}$ with
the standard univariate analysis. At the end of this procedure, we have 
full analytic expressions for the $g_{ij}^{\pm}$, and only
the polynomials $P_{ij}^\pm$ remain to be determined.

\subsection{Vandermonde Sampling Procedure}
\label{sec:VandermondeSampling}

Having determined the total degree of the $P_{ij}^\pm$ of \cref{eq:PartialFractionDecomposition}
and given that they depend on only five variables, 
we observe that the $P_{ij}^\pm$ have considerably fewer free Ansatz parameters than the
$n_i^\pm$ of \cref{eq:CommonDenominatorForm}. We have thus already taken an important step in 
addressing the problem identified at the end of section \cref{sec:cdAnsatz}.
In this section we describe a sampling procedure which allows for an efficient 
determination of the $P_{ij}^{\pm}$.
Specifically, we will use the Vandermonde-based approach introduced in 
ref.~\cite{Klappert:2019emp} in the context of univariate polynomial reconstruction
(see also \cite{Ellis:2007br} for an application of related ideas in the context
of one-loop amplitudes).
In our approach, which we will now detail, the 
generation of the numerical data is decoupled from the determination of the $P_{ij}^{\pm}$ using that data,
as is commonly the case for dense functional reconstruction algorithms (see e.g.\ \cite{Peraro:2016wsq}).

In order to determine the $P_{ij}^{\pm}(\vec{s}_{\mathrm{rem}})$, we must first discuss how to numerically
evaluate them. This is not trivial as we only have indirect access to them through \cref{eq:PartialFractionDecomposition}.
More explicitly, given some value $\vec{s}_{\mathrm{rem}}^{(k)}$ we cannot directly
evaluate the $P_{ij}^{\pm}(\vec{s}_{\mathrm{rem}}^{(k)})$ because we only have access
to the numerical value of the $r_i^\pm$
(using the algorithm outlined in \cref{sec:NumericalCalculation})
and the numerical value of the $g_{ij}^{\pm}$
(using their analytic expression). We can
however extract the value of the $P_{ij}^{\pm}(\vec{s}_{\mathrm{rem}}^{(k)})$
by sampling \cref{eq:PartialFractionDecomposition} over enough values of $s_{34}$,
that is by solving the system
\begin{equation}\label{eq:Peval}
  \left(
    \begin{matrix}
      g_{i 1}^{\pm}(\vec{s}_{\mathrm{rem}}^{(k)}, s_{34}^{(k, 1)}) & \cdots & g_{i n_i}^{\pm}(\vec{s}_{\mathrm{rem}}^{(k)}, s_{34}^{(k, 1)}) \\
      \vdots & \ddots & \vdots \\
      g_{i 1}^{\pm}(\vec{s}^{(k)}_{\mathrm{rem}}, s_{34}^{(k, n_i)}) & \cdots & g_{i n_i}^{\pm}(\vec{s}^{(k)}_{\mathrm{rem}}, s_{34}^{(k, n_i)}) \\
    \end{matrix}
  \right)
  \left( 
    \begin{matrix}
      P_{i 1}^{\pm}(\vec{s}_{\mathrm{rem}}^{(k)}) \\
      \vdots \\
      P_{i n_i}^{\pm}(\vec{s}_{\mathrm{rem}}^{(k)}) \\
    \end{matrix}
  \right)
=
  \left( 
  \begin{matrix}
    r_i^\pm(\vec{s}_{\mathrm{rem}}^{(k)}, s_{34}^{(k, 1)}) \\
    \vdots \\
    r_i^\pm(\vec{s}_{\mathrm{rem}}^{(k)}, s_{34}^{(k, n_i)}) \\
  \end{matrix}
\right),
\end{equation}
where  $s_{34}^{(k, i)}$ denotes the different values of $s_{34}$ that
are sampled (the different values are labelled by $i$, and $k$ is used 
to encode  the fact that these different values are associated with 
$\vec{s}_{\mathrm{rem}}^{(k)}$). The sampling is done in a finite field,
and the number of sample points, denoted by $n_i$, is the number of
terms in the partial-fraction decomposition Ansatz 
of~\cref{eq:PartialFractionDecomposition}. In practice, this number
is $\mathcal{O}(10)$.
We note that because we sample over arbitrary values of $s_{34}^{(k, i)}$,
it is easy to pick values such that $\mathrm{tr}_5^2$ is a
perfect square. Indeed, we recall that if one chooses points
at random in the finite field this will be the case 
50\% of the time \cite{Abreu:2020jxa, hardy1979introduction}.

Now that we have discussed how to numerically evaluate the 
$P_{ij}^{\pm}(\vec{s}_{\mathrm{rem}}^{(k)})$ for a given 
value $\vec{s}_{\mathrm{rem}}^{(k)}$, we describe how to use
the numerical evaluations to determine the analytic form of 
the $P_{ij}^{\pm}$. In principle we could sample the polynomials
on enough random values of $\vec{s}_{\mathrm{rem}}^{(k)}$, but this
would lead to unstructured linear systems, requiring the use of Gaussian elimination 
and creating a bottleneck in the procedure. 
Instead,  we sample them on values that guarantee that the linear system
to be solved corresponds to a (generalized) Vandermonde matrix.
In practice, we make use of the ideas introduced in 
ref.~\cite{Klappert:2019emp}, to which we refer for further details.
Here we simply outline the main steps of the approach.
We begin by considering a polynomial
\begin{equation}
  q(\vec{s}_{\mathrm{rem}}) = \sum_{\vec{\alpha}_i \in S} c_{\vec{\alpha}_i} m_{\vec{\alpha}_i}, \quad \mathrm{where} \quad m_{\vec{\alpha}_i} = s_{\sv 1}^{\alpha_{i,1}} s_{12}^{\alpha_{i,2}} s_{23}^{\alpha_{i,3}} s_{4\sv}^{\alpha_{i,4}} p_\sv^{2\alpha_{i,5}} ,
\end{equation}
and the sum over exponent vectors $\vec{\alpha}_i$ runs over some finite set 
$S$.
In the absence of further information on the polynomial, this set is
taken to be the full set of exponents of a homogeneous polynomial of fixed
degree (we recall that we have already determined the
degree of the $P_{ij}^{\pm}$).
We then introduce a so-called \emph{anchor point},
\begin{equation}
  \vec{s}_{\mathrm{rem}, (0)} = \left(s_{\sv 1, (0)}, s_{12, (0)}, s_{23, (0)}, s_{4\sv, (0)}, p_{\sv,(0)}^2 \right),
\end{equation}
from which we define further values $\vec{s}_{\mathrm{rem}}^{(k)}$ as
\begin{equation}
  \vec{s}_{\mathrm{rem}}^{(k)} = \left(s_{\sv 1, (0)}^{k}, s_{12, (0)}^{k}, s_{23, (0)}^{k}, s_{4\sv, (0)}^{k}, p_{\sv,(0)}^{2k} \right). 
\end{equation}
That is, the $k^\textrm{th}$ point is computed by taking the 
$k^\textrm{th}$ power of the  anchor point $\vec{s}_{\mathrm{rem}, (0)}$. 
The importance of this strategy is that the
monomials behave in a structured way. More precisely, it is not hard to see that
\begin{equation}
  m_{\vec\alpha_i}(\vec{s}_{\mathrm{rem}}^{(k)}) = \left[ m_{\vec\alpha_i}(\vec{s}_{\mathrm{rem}, (0)}) \right]^k.
\end{equation}
If we now evaluate the polynomial $q$ on $\vec{s}_{\mathrm{rem}}^{(k)}$
for $k = 1, \ldots, |S|$, then we end up with a constraining linear system for
the $c_{\vec{\alpha}_i}$ given by
\begin{equation}
  \left(
    \begin{matrix}
      \left[m_{\vec{\alpha}_1}(\vec{s}_{\mathrm{rem},(0)})\right]^1 & \cdots & \left[m_{\vec{\alpha}_{|S|}}(\vec{s}_{\mathrm{rem},(0)})\right]^1 \\
      \vdots & \ddots & \vdots \\
      \left[m_{\vec{\alpha}_1}(\vec{s}_{\mathrm{rem},(0)})\right]^{|S|} & \cdots & \left[m_{\vec{\alpha}_{|S|}}(\vec{s}_{\mathrm{rem},(0)})\right]^{|S|} \\
    \end{matrix}
  \right)
  \left( 
  \begin{matrix}
    c_{\vec{\alpha}_1}\\
    \ldots\\
    c_{\vec{\alpha}_{|S|}}\\
  \end{matrix}
\right)
  =
  \left(
    \begin{matrix}
      q(\vec{s}_{\mathrm{rem}}^{(1)}) \\
      \ldots \\
      q(\vec{s}_{\mathrm{rem}}^{(|S|)}) 
    \end{matrix}
  \right).
  \label{eq:VandermondeConstraint}
\end{equation}
The matrix in \cref{eq:VandermondeConstraint} is known as a
(generalized) Vandermonde matrix.
The special structure of this system allows it to be efficiently solved in
$\mathcal{O}(|S|^2)$ time and $\mathcal{O}(|S|)$ space. In practice, Vandermonde
systems with a side length $|S|$ of around $10^5$ can be solved
in just over a minute on a modern laptop computer.
Discussion of an efficient algorithm for solving the Vandermonde system can be
found in refs.~\cite{Press2007NumericalRT,Klappert:2019emp,PageSAGEXLectures}. 
We note that the Vandermonde matrix becomes singular if any of the elements 
of a row are identical, but we did not find this to occur in practice.

Equations \eqref{eq:Peval} and \eqref{eq:VandermondeConstraint} are 
the cornerstones of our reconstruction procedure. 
The first allows us to obtain numerical values for the polynomials $P_{ij}^\pm$,
and the second to use these numerical
values to determine the analytic form of the polynomials $P_{ij}^\pm$.
We find this reconstruction approach to have
several important benefits.
First, we note that it is common that, during the reconstruction procedure,
one of the $P_{ij}^{\pm}$ becomes completely constrained before the
full set of $P_{ij}^{\pm}$ is known. In our approach, known polynomials can
easily be removed in a procedure known as 
``pruning''~\cite{Klappert:2019emp}. This leads to a significant 
efficiency gain. Another advantage of our approach is that it is simple 
to generate a sufficient set of values
$\vec{s}_{\mathrm{rem}}^{(k)}$  and $s_{34}^{(k, i)}$ required for the
reconstruction of all $P_{ij}^{\pm}$, and a cluster can then be used to evaluate the associated
$r_i^{\pm}$, which is the most expensive part of the procedure.
Furthermore, the approach is  agnostic to the specifics of
the set $|S|$. 
In practice, when one has performed the calculation over the first
finite field, one notices that many of the coefficients $c_{\vec{\alpha_i}}$ are
zero. With our strategy, when further finite fields are required 
it is trivial to implement zero coefficient constraints by simply repeating the
procedure with a reduced set $S$, leading to further efficiency gains. 
Finally, we remark that the result is
automatically homogeneous, and no procedure of de-homogenizing and
re-homogenizing is required.


\section{Implementation and Results}
\label{sec:Results}

\subsection{Implementation}

The numerical reduction of the amplitudes to master integrals is performed
with the two-loop numerical unitarity approach, as implemented 
in \Caravel{}~\cite{Abreu:2020xvt}. \Caravel{} was updated with the required
Feynman rules, and the surface terms required for five-point one-mass planar
two-loop topologies, as discussed in \cref{sec:NumericalCalculation}.
Starting from the decomposition into master integrals, 
where we use the basis of master integrals of ref.~\cite{Abreu:2020jxa}, we then obtain
the decomposition in terms of the pentagon functions of ref.~\cite{Chicherin:2021dyp}.
This allows us to obtain the decomposition of the remainders of \cref{eq:pentFuncFF} 
with numerical coefficients.

These numerical evaluations are then used to obtain analytic expressions
for the coefficients $r_i(\vec s,\trFive)$ of \cref{eq:pentFuncFF} with
the procedure described in \cref{sec:reconstruction}.
In order to perform the univariate and bivariate reconstructions necessary for
establishing the partial-fraction Ansatz of \cref{eq:PartialFractionDecomposition},
we use an in-house \texttt{C++} implementations of Thiele's algorithm and of
the algorithm of ref.~\cite{Peraro:2016wsq}, respectively. 
We then use \texttt{Mathematica} to perform the partial
fractioning of the bivariate functions with respect to $s_{34}$.
To simplify the reconstruction procedure, we 
work with only a linearly-independent set of the rational functions present in the
remainder \cite{Abreu:2018aqd,Abreu:2019rpt}. 
That is, we express each (form factor) remainder $\mathcal{R}^{\{i\}}$, 
see \cref{eq:pentFuncFF}, in the form
\begin{equation}\label{eq:results-format}
  \mathcal{R}^{\{i\}} = \sum_{k \in K, b \in B} \tilde{r}_k M_{kb} h_b ,
\end{equation}
where $M_{kb}$ is a matrix of rational numbers, and $K$ indexes the basis of
rational functions $\tilde{r}_k$ of the corresponding remainder.
The $\tilde{r}_k$ are reconstructed with the approach described in
\cref{sec:VandermondeSampling}, which allows for an efficient solution of the
required linear systems since they are in Vandermonde form.
This procedure was implemented in a combination of \texttt{Mathematica} and
Rust routines. Specifically, the solution of the Vandermonde system itself was
implemented in Rust and linked into the \texttt{Mathematica} program via
\texttt{LibraryLink}.

To illustrate the complexity of the reconstruction procedure and the impact
of the partial-fraction Ansatz over the common denominator Ansatz, we
record the number of free coefficients in the two Ans\"atze in 
\cref{tab:QuarkResultsData,tab:GluonResultsData}. 
In these tables we do not consider $N_f^2$, as those functions are trivial
compared to the $N_f^0$ and $N_f^1$ contributions.
It can clearly be seen that, for all amplitudes, the univariate partial
fractioning has a large effect, reducing the dimension of the Ansatz
by a factor of up to $\sim 50$.
We note that even after partial fractioning there are still contributions
with $\mathcal{O}(500\,\textrm{k})$ undetermined parameters,  
and there is a large number of linear systems with $\mathcal{O}(100\,\textrm{k})$
side length to be solved. By having them in Vandermonde form, the solutions
can be obtained in $\mathcal{O}(1\,\textrm{min})$ on a laptop.
Furthermore, we find that once the Ansatz has
been fit, there is a further strong reduction in the size of the
result as many of the coefficients turn out to be zero (see the last column
of \cref{tab:QuarkResultsData,tab:GluonResultsData}). Given this sparsity, it
would be interesting to consider applying sparse algorithms such as 
the Ben-Or/Tiwari algorithm~\cite{ben1988deterministic, Klappert:2020aqs}.
The fact that so many coefficients are zero has another important consequence
on the efficiency of our approach, as the set of contributing monomials is
known after performing the computation in the first finite field.
The Vandermonde-based procedure outlined in \cref{sec:VandermondeSampling}
is particularly suited to make use of the smaller basis of monomials when computing
in more finite fields.
In practice, only two finite fields of cardinality $\mathcal{O}(2^{31})$ were
required to perform the rational reconstruction.

\begin{table}
  \renewcommand{\arraystretch}{1.2}
  \begin{tabular}{cccccc}
    \toprule
    &&& \multicolumn{2}{c}{Max Ansatz Size}   & Max Non-Zero Terms  \\
  $\mathcal{R}_\gluon$ & $p_5 \parallel p_i$ & |K| & Common Denominator & Partial Fractioning & Result \\
    \midrule
  \multirow{3}{*}{$++$ $N_f^0$} &1 & 29 & $\num{660  }$\,k & $\num{58 }$\,k & $\num{9  }$\,k \\
   &2 & 32 & $\num{1100 }$\,k & $\num{82 }$\,k & $\num{14 }$\,k \\
   &3 & 32 & $\num{960  }$\,k & $\num{87 }$\,k & $\num{15 }$\,k \\
    \hline 
    \multirow{3}{*}{$++$ $N_f^1$} &1 & 23 & $\num{380  }$\,k & $\num{28 }$\,k & $\num{5.2  }$\,k \\
   &2 & 23 & $\num{750  }$\,k & $\num{54 }$\,k & $\num{8  }$\,k \\
   &3 & 23 & $\num{580  }$\,k & $\num{63 }$\,k & $\num{11 }$\,k \\
\hline              
    \multirow{3}{*}{$+-$ $N_f^0$} &1 &58& $\num{5500 }$\,k & $\num{180}$\,k & $\num{37 }$\,k \\
   &2 &67& $\num{7000 }$\,k & $\num{480}$\,k & $\num{110}$\,k \\
   &3 &67& $\num{5900 }$\,k & $\num{380}$\,k & $\num{90 }$\,k \\
    \hline          
    \multirow{3}{*}{$+-$ $N_f^1$} &1 &50& $\num{4600 }$\,k & $\num{160}$\,k & $\num{33 }$\,k \\
  &2 &53& $\num{5000 }$\,k & $\num{380}$\,k & $\num{87 }$\,k \\
   &3 &53& $\num{4200 }$\,k & $\num{310}$\,k & $\num{75 }$\,k \\
    \hline          
    \multirow{3}{*}{$-+$ $N_f^0$} &1 &75& $\num{12000}$\,k & $\num{210}$\,k & $\num{46 }$\,k \\
   &2 &85& $\num{14000}$\,k & $\num{500}$\,k & $\num{130}$\,k \\
   &3 &85& $\num{24000}$\,k & $\num{430}$\,k & $\num{99 }$\,k \\
    \hline          
    \multirow{3}{*}{$-+$ $N_f^1$} &1 &44 & $\num{4600 }$\,k & $\num{120}$\,k & $\num{25 }$\,k \\
   &2 &49 & $\num{3800 }$\,k & $\num{210}$\,k & $\num{54 }$\,k \\
   &3 &49 & $\num{8900 }$\,k & $\num{270}$\,k & $\num{63 }$\,k \\
    \hline
  \end{tabular}
  \caption{
Characterizing information for remainders with $\kappa = \gluon$ at various stages of the
    computation. 
We specify the remainders by the helicity
    states of the gluon pair and the power of $N_f$ (the $N_f^2$ contributions
    are trivial compared to the other powers, so we do not list them).
     $|K|$ is the
    dimension of the space of rational functions of the corresponding amplitude.
   The maximal Ansatz size in common denominator and partial-fraction form is given, when considered 
   over all rational functions. 
   The last column, `Max Non-Zero Terms',  gives the largest number of non-zero terms in the 
   result, again taken over the basis of rational functions. 
    Term counts are given to two significant digits for readability. }
  \label{tab:GluonResultsData}
\end{table}

\begin{table}
  \renewcommand{\arraystretch}{1.2}
  \begin{tabular}{cccccc}
    \toprule
   &&& \multicolumn{2}{c}{Max Ansatz Size}   & Max Non-Zero Terms   \\
  $\mathcal{R}_\Quark$ & $p_5 \parallel p_i$ & |K| & Common Denominator & Partial Fractioning & Result  \\    
    \midrule
      \multirow{3}{*}{$+-$ $N_f^0$} &1& 50  &$\num{1200} $\,k & $\num{53}$\,k& $\num{11}$\,k \\
       &2& 57 &$\num{1700}$\,k & $\num{210} $\,k& $\num{56} $\,k \\
       &3& 56 &$\num{1400}$\,k & $\num{240} $\,k& $\num{56} $\,k \\
    \hline               
      \multirow{3}{*}{$+-$ $N_f^1$} &1& 18 &$\num{26}$\,k & $\num{13}$\,k& $\num{1.5}$\,k\\
       &2& 20 &$\num{140 }$\,k & $\num{47}$\,k& $\num{5.1}$\,k\\
       &3& 20 &$\num{140} $\,k & $\num{54} $\,k& $\num{8.4}$\,k\\
    \hline               
    \multirow{3}{*}{$-+$ $N_f^0$} &1& 69 &$\num{2300} $\,k & $\num{64} $\,k& $\num{14} $\,k \\
       &2& 75 &$\num{2300}$\,k & $\num{230} $\,k& $\num{57} $\,k \\
       &3& 79 &$\num{5500}$\,k & $\num{220}$\,k& $\num{48}$\,k \\
    \hline               
    \multirow{3}{*}{$-+$ $N_f^1$} &1& 30 &$\num{240}$\,k & $\num{21}$\,k& $\num{3.9}$\,k\\
       &2& 31 &$\num{380}$\,k & $\num{52}$\,k& $\num{11}$\,k \\
       &3& 31  &$\num{380}$\,k & $\num{51}$\,k& $\num{9.1}$\,k\\
    \hline
  \end{tabular}
  \caption{
    Characterizing information for remainders with $\kappa = \Quark$ at
    various stages of the computation. We specify the remainders by the helicity states of the quark
    pair not coupled to the vector boson, and the power of $N_f$ (the $N_f^2$ contributions
    are trivial compared to the other powers, so we do not list them). Column headings are identical to table
\ref{tab:GluonResultsData}. Term counts are given to two significant digits for readability.}
  \label{tab:QuarkResultsData}
\end{table}

\subsection{Results and Validation}
\label{sec:results-validation}

The main result of this paper are the analytic expressions for the
two-loop remainders $\mathcal{R}_\kappa^{(2)[j]}$. They are given
in a set of ancillary files that can be obtained from \cite{W4partonsite}. 
Using the definition of the remainders in \cref{eq:remainder2l}, 
we can also assemble the two-loop amplitudes  $\CA^{(2)[j]}$. This requires
the knowledge of the one-loop amplitudes $\CA^{(1)[j]}$, and we have thus also 
recomputed them using similar reconstruction techniques. In fact, the one-loop
amplitudes are presented as a decomposition in terms of master integrals,
that is we reconstruct the coefficients $c_{\Gamma,i}$ in the one-loop
equivalent of the decomposition of \cref{eq:A}, where for the basis
of master integrals we take the one-loop basis used in \cite{Abreu:2020jxa}.
We also include a map from the one-loop integrals to the pentagon functions
of ref.~\cite{Chicherin:2021dyp} up to weight 4, 
so that the one-loop expressions can be written as a decomposition in terms
of pentagon functions similar to the format we use for the two-loop remainders.
Provided the one-loop integrals are known, the expressions we present for the one-loop amplitudes 
can however be used to expand the amplitudes to arbitrary order in $\epsilon$,
extending the results of refs.~\cite{Bern:1997sc, Bern:1996ka}.
All the components required to assemble the one-loop amplitudes are
also in the set of ancillary files.

To facilitate the use of our ancillary files, we include a \texttt{Mathematica} script
called \texttt{amp\_eval.m} which assembles the different components required to
evaluate the one-loop and two-loop amplitudes and remainders, and evaluates
them at the phase-space point specified in \cref{sec:benchmark}.
The target values obtained by running this script are also included in \cref{sec:benchmark},
see \cref{tab:values-bare-amplitudes-1l,tab:values-bare-amplitudes-2l,tab:values-finite-remainders}.
Together with the \texttt{README.md} file, the script was prepared to
document the different files and be a good entry point to start 
exploring our results. 

Let us briefly comment on the analytic structure of the two-loop amplitudes
we computed.
It was observed in ref.~\cite{Badger:2021nhg} that 
the pentagon functions involving the letters $\{W_{16},W_{17},W_{27},W_{28},W_{29},W_{30}\}$  
(we use the notation of ref.~\cite{Abreu:2020jxa}), 
which are present in the contributing master integrals,
drop out from the squared amplitudes for on-shell $Wb\overline{b}$ 
production when expanded to finite order in $\epsilon$.
The same observation was made in ref.~\cite{Badger:2021ega} for the two-loop finite
remainders for the production of a $H$ boson in association with a $b\bar{b}$ pair.
We observe here that the same holds also at the level of helicity form factors 
for both the quark and gluon partonic processes considered in this work.
Furthermore, we also observe that the pentagon functions involving the letter $\mathrm{tr}_5$
do not appear in the finite remainders, an interesting
fact that has been previously linked to cluster algebras~\cite{Chicherin:2020umh}.
We note that the pentagon-function basis of ref.~\cite{Chicherin:2021dyp} has been 
constructed in a way that such cancellations are manifest.

We have performed a number of checks, not only on our final results, but also
on the intermediate steps of our calculation.
Let us first discuss the internal checks performed on the intermediate
stages of the calculation.
Since this is the first time we have used the surface terms for five-point
one-mass kinematics (see the discussion in \cref{sec:2loopNumUni}), we have 
numerically cross-checked them against \texttt{FIRE}~\cite{Smirnov:2008iw,Smirnov:2014hma}.
The numerical evaluation of the amplitudes was performed within the 
numerical-unitarity code \Caravel{}~\cite{Abreu:2020xvt},
which includes many internal self-consistency checks.
The numerical calculation of the two-loop remainders is also performed within
\Caravel{}, and at each phase-space point we thus verify that the amplitude
has the pole structure predicted by \cref{eq:remainder2l}.
After obtaining the analytic expression for the one-loop amplitudes
and the two-loop remainders, we performed consistency checks to verify
that the reconstructed expressions agree with the numerical evaluations obtained
with \Caravel{}. Finally, we verified that the reconstructed expressions,
which are obtained from numerical evaluations in a finite field, agree
with floating-point numerical evaluations within \Caravel{}.

Let us now discuss the checks we have made on the final results we have obtained.
First, the one-loop amplitudes were checked up to order $\epsilon^0$ with the results
obtained from the \texttt{BlackHat} library~\cite{Berger:2008sj}. 
Second, we reproduced the numerical table of ref.~\cite{Hartanto:2019uvl} using 
\Caravel{} to evaluate the master integral coefficients and 
\texttt{DiffExp}~\cite{Hidding:2020ytt} to evaluate the master integrals.
This is a strong check of the correctness of the numerical evaluations within 
\Caravel{}, which we used to obtain our analytic expressions.
Third, we reproduced the results of ref.~\cite{Badger:2021nhg} for unpolarized on-shell $Wb\overline{b}$ production 
by squaring our currents $A^{\mu}$ in \cref{eq:CurrentFactorization} appropriately.\footnote{More specifically, we agree
with the revised version of the results presented in 
ref.~\cite{Badger:2021nhg}.}
This is a particularly stringent test, as the calculation in ref.~\cite{Badger:2021nhg} is performed in the
conventional dimensional regularization (CDR) scheme, and using the Larin's 
$\gamma_5$ prescription \cite{Larin:1993tq}. The latter produces
non-trivial differences between vector and axial currents, and
we find agreement at the level of the finite remainders. 

Finally, let us mention that given our analytic results in the form of \cref{eq:results-format}, which are valid for the partonic channels 
of \cref{eq:partProcesses} with momenta $p_1$ and $p_2$ incoming, it is straightforward to obtain results in other partonic channels by permutations of particles' momenta.
Indeed, the action of permutations on the rational functions $\tilde{r}_i(\vec{s}, \mathrm{tr}_5)$ is obvious.
The action of permutations on the one-mass pentagon functions $h_i$ is discussed in section 3.2 of ref.~\cite{Chicherin:2021dyp},
and is explicitly provided in the supplementary materials thereof. 
It is worth noting that the latter allows us to replace the dedicated analytic continuation procedure which had to be employed, for instance, in ref.~\cite{Abreu:2021oya},
by simple substitutions.


\section{Conclusions and Outlook}
\label{sec:Conclusions}

We have computed the analytic expressions for the planar two-loop
QCD corrections to the helicity amplitudes for four partons
and a vector boson that decays into a lepton pair.
These expressions allow us to compute the full leading color
two-loop QCD corrections to amplitudes for four partons
and a $W$ boson that decays into a lepton pair. Furthermore, 
they also fully determine the gauge-invariant planar contributions to the
same amplitudes where the $W$ boson is replaced by a $Z/\gamma^*$
(in this case, some leading-color corrections have non-planar
contributions, but they can be separated from the contributions
we compute because they have a distinct coupling structure).

The analytic results we present are reconstructed from numerical
evaluations, an approach that has been very successfully applied to
five-point massless processes. We find that there is a marked
increase in complexity when considering five-point one-mass kinematics.
This led us to develop a more efficient reconstruction approach,
building on a partial-fraction Ansatz \cite{Badger:2021nhg} and
on a judicious numerical sampling procedure that leads to 
linear systems in Vandermonde form. Conveniently, the numerical evaluations of
the amplitudes and the reconstruction of the analytic functions
remain decoupled in this approach.

We have explicitly presented the analytic expressions that are valid in the region where $p_1$ and $p_2$ are incoming. 
The expressions in the other regions can be straightforwardly obtained by considering appropriate combinations of particles' momenta permutations, charge and parity conjugation.
Given the analytic complexity of the derived amplitudes and the large number of required partonic channels, we leave an efficient implementation of their numerical evaluation and the study of numerical stability for future work.

Given the remarkable progress in handling the
real-radiation contribution with complex final-state jet structure at NNLO QCD
(see e.g.~\cite{Czakon:2021mjy}), we expect that NNLO QCD predictions for the production of
a $W$ boson in association with two jets at hadron colliders are now within reach.

\begin{acknowledgments}
We thank Simone Zoia for communication on the comparison with the results of ref.~\cite{Badger:2021nhg},
and for discussions on the sign of parity-odd pentagon functions.
This project has received funding from the European's Union Horizon 2020
research and innovation programmes \textit{LoopAnsatz} (grant agreement number 896690)
and \textit{Novel structures in scattering amplitudes} (grant agreement number 725110).
M.K.'s work is funded by the German Research Foundation (DFG)
within the Research Training Group GRK 2044.
The work of F.F.C.\ is supported in part by the U.S.\ Department of Energy under
grant DE-SC0010102.
This work was performed on the bwUniCluster funded by the Ministry of Science,
Research and the Arts Baden-W\"urttemberg and the Universities of the State of
Baden-W\"urttemberg, Germany, within the framework program bwHP.  
Some of the computing for this project was performed on the HPC cluster at the
Research Computing Center at the Florida State University (FSU).
\end{acknowledgments}

\appendix

\section{Axial-vector and Vector Couplings}
\label{sec:gamma5}

In this section, we discuss the equivalence of  vector and axial-vector currents  up to a sign in massless QCD,
which holds as long as the axial-vector couplings to closed fermion loops are excluded.

\begin{figure}[t]
  \centering
  \includegraphics[width=0.3\textwidth]{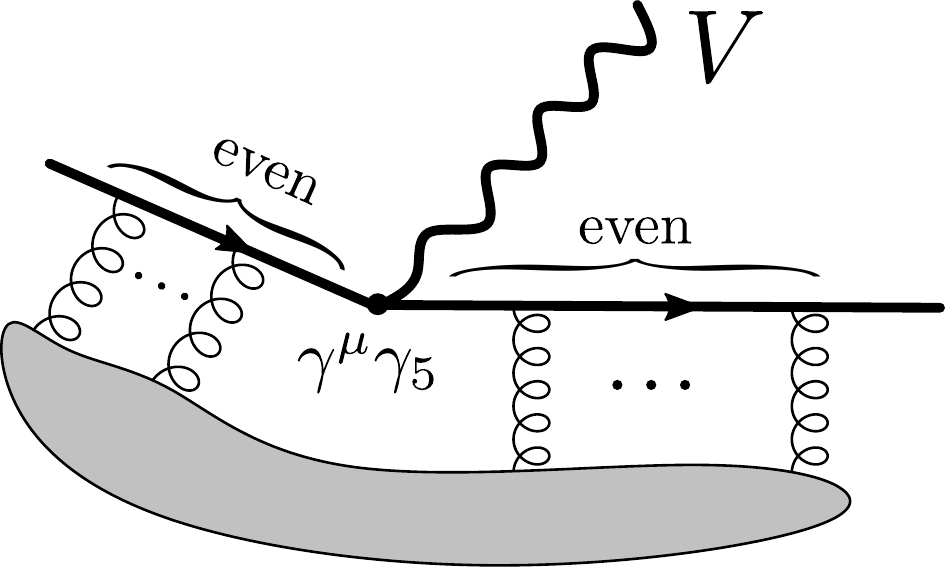}
  \caption{
    The structure of diagrams contributing to axial-vector currents considered in this work.
    The shaded blob contains the remaining external particles, as well as loops.
    Each gluon that couples to the quark line adds two additional $\gamma$-matrices to the spinor chain, 
    one from the interaction vertex and one from the additional quark propagator.
  }
  \label{fig:spchain}
\end{figure}

The diagrams contributing to the vector and axial-vector current are schematically represented in \cref{fig:spchain}.
In four dimensions these diagrams give rise to the spinor chains
\begin{subequations}\label{eq:sp-chain-4d}
  \begin{align}
    S^\mu & =  \bar u^{R}\, \cdots{} \gamma^\mu \cdots{} \, v^L, \\
    S^\mu_5 & =   \bar u^{R}\, \cdots{} \gamma^\mu\gamma^5  \cdots{} \, v^L\,,
  \end{align}
\end{subequations}
where we use lower case $\gamma's$ for the four dimensional Clifford algebra and 
$\gamma^5:=\tilde \gamma= \ii \gamma^0\gamma^1\gamma^2\gamma^3$ with $(\gamma^5)^2=\mathbb{1}$. 
Without loss of generality, 
we consider the spinor chains with left-handed quarks and right-handed anti-quarks.

We begin by confirming that in four dimensions the currents differ by sign,
\begin{equation}\label{eq:currentRel}
 S^{\mu} = - S^{\mu}_5\,.
\end{equation}
Indeed, by inspecting the diagrams in \cref{fig:spchain} we observe
that we can always anticommute $\gamma^5$ through an even number of $\gamma$-matrices to 
the right (or odd number of $\gamma$ matrices to the left) and use the definition,
\begin{equation}\label{eq:g5}
  \gamma^5  \, v^L = -v^L \quad \mbox{and} \quad \bar u^{R}(p)\,\gamma^5 =  \bar u^{R}(p)\,,
\end{equation}
to eliminate $\gamma^5$ from any spinor chain.

In the following we discuss that the same relation holds in dimensional regularization 
within Kreimer's anticommuting $\gamma^5$ scheme \cite{Kreimer:1989ke,Korner:1991sx}.
In this scheme one starts with the $D$-dimensional Clifford algebra 
$\{ \Gamma^\mu\}$,
\begin{equation}\label{eq:ddim-cl-algebra}
  \left\{\Gamma^\mu,\Gamma^\nu\right\} =  2 g^{\mu\nu}\,\mathbb{1}\,, 
  \quad \left\{\tilde \Gamma, \Gamma^\mu\right\} = 0\quad \mbox{and} 
  \quad \tilde\Gamma^2 = \mathbb{1}.
\end{equation}
Here $\tilde\Gamma$ is the $D$-dimensional generalisation of $\gamma^5$, which allows to 
construct chirality projectors. We will also use the literal embedding of $\gamma^5$ in the D-dimensional Clifford algebra, namely
\begin{equation}\label{eq:G5}
  \Gamma^5= \ii \Gamma^0\Gamma^1\Gamma^2\Gamma^3 \,.
\end{equation}
Any element $F $ in the $D$-dimensional Clifford algebra can be expanded into a basis 
\begin{equation}\label{eq:cl-algebra-basis} 
  F = f \mathbb{1} + f_{\mu} \Gamma^{\mu} + f_{\mu_1\mu_2} \Gamma^{\mu_1\mu_2} + f_{\mu_1\mu_2\mu_3} \Gamma^{\mu_1\mu_2\mu_3}
     + \ldots{},
\end{equation}
where the basis elements $\Gamma^{\mu_1\cdots{}\mu_n}$  are anti-symmetrized products of $n$ $\Gamma$-matrices. 
In Kreimer's $\gamma^5$ scheme one replaces 
$\gamma^{\mu}\to \Gamma^\mu, ~ \gamma^5 \to \tilde\Gamma$,
and the anticommutativity of $\tilde\Gamma$ is used to ensure that at most one 
$\tilde\Gamma$ appears in each spinor chain.
The evaluation of spinor chains is formulated in terms of trace operations $\Tr[\cdot]$, which return 
particular coefficients in the decomposition of \cref{eq:cl-algebra-basis},
\begin{subequations}\label{eq:kreimer-traces}
  \begin{align}
    \Tr[F] & \coloneqq  4\, f = 4\, \frac{\tr[F]}{\tr[\mathbb{1}]}, \\
    \Tr[\tilde \Gamma F] & \coloneqq  4\,f_{0123} =  4\,\frac{ \tr[\Gamma^5 F]}{\tr[\mathbb{1}]}\,.
  \end{align}
\end{subequations}
Here $\tr[\cdot]$ is the conventional trace for $\Gamma$-matrices. 
With this definition, 
a single $\tilde \Gamma$, if present, is moved to the left in products of $\Gamma$-matrices. This operation is always 
possible in fixed-order computations based on Feynman rules.
In the second line we used the orthogonality of the basis $\Gamma^{\mu_1 … \mu_n}$ to 
obtain the coefficient $f_{0123}$ using the insertion $\Gamma^5$ in $\tr[\cdot]$ \cite{Korner:1991sx}. 
In practice, we can work without making the $\tilde\Gamma$ factor manifest and obtain the convenient 
trace prescription,
\begin{equation}
  F'=\tilde \Gamma F \,,\quad  \Tr[ F']  =   \Tr[ \tilde \Gamma^2 F' ] =  \Tr[ \tilde \Gamma (\tilde \Gamma F')] =
    4\,\frac{ \tr[ \Gamma^5\tilde\Gamma  F']}{\tr[\mathbb{1}]}\,.
\end{equation}
This amounts to inserting $( \Gamma^5\tilde\Gamma)$ at a fixed point in the spinor chain before evaluating the conventional trace $\tr[\cdot]$. 
As we will now see, for spinor chains with external states the insertion point is naturally given.

We are now in the position to discuss the effect of the above prescription for evaluating the 
spinor chains $S^\mu$ and $S^\mu_5$.
First we use the $D$-dimensional generalization of the spinor states
discussed in ref.~\cite{Abreu:2018jgq},
which are a tensor product of four-dimensional states and a spinor basis beyond four dimensions. 
For simplicity we denote two such spinors by the same symbols as above $\bar u^{R}$ and $v^L$ 
and use $\bar u^{R} \Gamma^5 = \bar u^{R}$ and $ \Gamma^5 v^L = - v^L$, 
which is a trivial consequence of \cref{eq:G5}.
We obtain,
\begin{subequations}\label{eq:sp-chain-d}
  \begin{align}
    S^{\mu} & \coloneqq  \bar u^{R}\, \cdots{} \Gamma^\mu \cdots{} \, v^{L}, \\
    S^{\mu}_5 & \coloneqq   \bar u^{R}\, \cdots{}  \Gamma^\mu\tilde\Gamma \cdots{} \, v^{L}\,.
  \end{align}
\end{subequations}
We now rewrite the spinor chains in terms of the $\Tr[\cdot]$ operation. For the contributions including $\tilde\Gamma$ 
we have to consistently chose an insertion point for $\Gamma^5\tilde\Gamma$. A well defined choice is to insert after 
the $\bar u$ spinor associated to the external state. 
Using the properties in \cref{eq:ddim-cl-algebra,eq:kreimer-traces} we obtain
\begin{align}\begin{split}
  S^\mu_5 & =  \Tr[ v^{L}\bar u^{R}  \cdots{} \Gamma^\mu \tilde\Gamma  \cdots{}] = 
  		\tr[ v^{L}\bar u^{R} (\Gamma^5 \tilde\Gamma)    \cdots{} \Gamma^\mu\tilde\Gamma   \cdots{}] \\
		& = -\tr[ v^{L}\bar u^{R} \Gamma^5    \cdots{} \Gamma^\mu \cdots{} ] = 
		- \tr[ v^{L}\bar u^{R}   \cdots{} \Gamma^\mu \cdots{} ]\\
		&  = - S^\mu \,.
\end{split}\end{align}
where we used that $\bar u^{R} \Gamma^5 = \bar u^{R}$ and $\{\Gamma^\mu,\tilde\Gamma\}=0$.
This demonstrates that $S^\mu=-S^\mu_5$ can be maintained in dimensional regularization.

As pointed out in refs.~\cite{Kreimer:1989ke,Korner:1991sx} the trace operation $\Tr[\cdot]$ involving $\tilde\Gamma$ does not satisfy cyclicity.
Generally, this implies that all traces must be consistently evaluated with reference to a point where we insert $\Gamma^5\tilde\Gamma$.
Conveniently, the spinor chains in \cref{eq:sp-chain-d} have a distinguished  insertion point, namely next to the $\bar u$ spinor.
This insertion point is not only unambiguous, but also aligns signs of the 
four-dimensional algebra and the $D$-dimensional one.
Finally, let us recall that we do not consider contributions with axial couplings to closed fermion loops,
therefore we are not concerned with axial anomalies in this paper.

In summary we showed the equivalence of vector and axial-vector currents for spinor chains bounded by external states. 
In addition we obtain a simple prescription to handle axial-vector couplings following refs.~\cite{Kreimer:1989ke,Korner:1991sx}. 
This prescription is particularly convenient for our approach to dimensional regularization of helicity amplitudes \cite{Abreu:2018jgq},
which is realized entirely in integer dimensions and therefore suitable for direct numerical implementation.


\section{Rationalization of $\mathrm{tr}_5$}
\label{sec:ratParam}

In this appendix we present the explicit form of the rational parametrization of the five-point
one-mass phase-space which we use in the procedure described in \cref{sec:partFracAns} 
to constrain the partial-fraction Ansatz of \cref{eq:PartialFractionDecomposition}.
To rationalize $\trFive$ we use a parametrization in which only one variable is not a
Mandelstam invariant. More explicitly, we use the variables
\begin{equation}
\{  s_{\sv 1}, s_{12}, s_{23} , s_{4\sv}, \pv^2 , x  \},
  \label{eq:RationalParameters}
\end{equation}
where $x$ is chosen such that  $s_{34}$ and $\mathrm{tr}_5$ are rational in the 
variables of \cref{eq:RationalParameters}. Explicitly, they are given by
\begin{align}
  \begin{split}
\mathrm{tr}_5( &s_{\sv 1}, s_{12}, s_{23} , s_{4\sv }, p_{\sv }^2 , x ) = 
                           \frac{s_{\sv 1}(s_{4\sv } - s_{12}) + (s_{12} - p_{\sv }^2 ) s_{23}}{
4\, \big((s_{4\sv } - s_{23})^2 - 
 [s_{\sv 1}(s_{4\sv } - s_{12}) + (s_{12} - p_{\sv }^2) s_{23}]^2x^2)\big)}\\
                         &\times \Big[(s_{4\sv } - s_{23})^2 + \\
               & \quad \,\, 2\big(s_{\sv 1}s_{4\sv }(s_{12} \!-\! s_{4\sv }) + 
                 [p_{\sv }^2(s_{4\sv } \!-\! 2s_{12}) + s_{4\sv }s_{12} + s_{\sv 1}(s_{4\sv } \!+\! s_{12})]s_{23}
                 - (p_{\sv }^2 \!+\! s_{12})s_{23}^2\big)x  \\
&\quad   + (s_{\sv 1}(s_{4\sv } - s_{12})
+ (s_{12} -p_{\sv }^2 )s_{23})^2x^2\Big]\,,
\end{split}
\\
  \begin{split}
s_{34}( &s_{\sv 1},  s_{12}, s_{23} , s_{4\sv }, p_{\sv }^2 , x ) =
          \frac{2}{(s_{\sv 1}(s_{4\sv } - s_{12}) + ( s_{12} - p_{\sv }^2)s_{23})^2x^2 - (s_{4\sv } - s_{23})^2 }  \\
          &\times \Big[( s_{\sv 1}s_{4\sv }s_{12} -s_{\sv 1}s_{4\sv }^2 + p_{\sv }^2s_{4\sv }s_{23}
+ s_{\sv 1}s_{4\sv }s_{23} - 2p_{\sv }^2s_{12}s_{23} +  s_{\sv 1}s_{12}s_{23} \\
& \quad + s_{4\sv }s_{12}s_{23} - p_{\sv }^2s_{23}^2 - s_{12}s_{23}^2 
+ (s_{\sv 1}(s_{4\sv } - s_{12}) + (s_{12} -p_{\sv }^2 )s_{23})^2x)\Big]\,.
\end{split}
\end{align}

In order to disentangle the $r_i^+(\vec s)$ and $r_i^-(\vec s)$ contributions in the 
coefficients $r_i(\vec s,\trFive)$  of
\cref{eq:parityDecomposition}, we need to identify the phase-space point that is related by parity conjugation
to the point corresponding to \cref{eq:RationalParameters}. Indeed, parity conjugation flips the sign of $\trFive$
and leaves the Mandelstam variables invariant, and by evaluating $r_i(\vec s,\trFive)$ in pairs of parity-conjugate points we can determine $r_i^+(\vec s)$ and $r_i^-(\vec s)$. The parity-conjugate point
\begin{equation}
\{  s_{\sv 1}, s_{12}, s_{23} , s_{4\sv}, \pv^2 , \bar{x}  \},
\end{equation}
must be such that
\begin{align}\begin{split}
  \mathrm{tr}_5(  s_{\sv 1}, s_{12}, s_{23} , s_{4\sv }, p_{\sv }^2 , \bar x ) = -\mathrm{tr}_5(s_{\sv 1}, s_{12}, s_{23} , s_{4\sv }, p_{\sv }^2 , x),  \\
  s_{34}( s_{\sv 1}, s_{12}, s_{23} , s_{4\sv }, p_{\sv }^2 , \bar x ) = s_{34}( s_{\sv 1}, s_{12}, s_{23} , s_{4\sv }, p_{\sv }^2 , x ).
\end{split}\end{align}
Solving for $\bar x$, we find
\begin{align}
  \begin{split}
\bar{x} &= \Big[ -(s_{4\sv } - s_{23})^2 + (s_{\sv 1}s_{4\sv }(s_{4\sv } - s_{12}) - (p_{\sv }^2(s_{4\sv }  - 2s_{12})  \\
&  \qquad + s_{4\sv }s_{12} + s_{\sv 1}(s_{4\sv } + s_{12}))s_{23} + (p_{\sv }^2 + s_{12})s_{23}^2)x \Big]/\Big[  s_{\sv 1}s_{4\sv }s_{12} \\
& \qquad  - s_{\sv 1}s_{4\sv }^2 + p_{\sv }^2s_{4\sv }s_{23} + s_{\sv 1}s_{4\sv }s_{23} - 2p_{\sv }^2s_{12}s_{23} + s_{\sv 1}s_{12}s_{23}  \\
&  \qquad + s_{4\sv }s_{12}s_{23}  - p_{\sv }^2s_{23}^2 - s_{12}s_{23}^2 + (s_{\sv 1}(s_{4\sv } - s_{12})  + (-p_{\sv }^2 + s_{12})s_{23})^2x \Big] \,.
  \end{split}
\end{align}


\section{Benchmark numerical evaluation}\label{sec:benchmark}

To facilitate comparisons with our results, we provide benchmark 
numerical evaluations for all the helicity amplitudes we computed. 
We evaluate the bare helicity amplitudes and finite remainders on a generic point 
from physical six-point massless phase-space, where the momenta $p_1$ and $p_2$ are incoming.
The point we choose corresponds to
\begin{equation}\label{eq:psPointNum}
  \begin{gathered}
    s_{12} = 5, \quad
    s_{23}  = -\frac{1}{3}, \quad
    s_{34}  = \frac{11}{13}, \quad
    s_{456} = \frac{17}{19}, \quad
    s_{156} = -\frac{23}{29}, \quad 
    s_{56} = \frac{1}{7}, \\
    s_{345}  = \frac{4304788}{896077}, \quad
    s_{45}  = \frac{2911673}{3928953}, \quad
    s_{16}  = -\frac{186065}{1998941}, \qquad
    \trFive = -\frac{10 \ii \sqrt{292010395}}{150423},
  \end{gathered}
\end{equation}
where the first line specifies the reduction to the five-point one-mass kinematics (we
recall that $p_\sv=p_5+p_6$).

We present the amplitudes for four partons and $W^+ (\to \bar{e}^+ \nu^-)$, i.e.\
\begin{align}\begin{split}\label{eq:amplitudesNum}
  \mathcal{A}^{(l)[j]}_{\gluon;h_2 h_3} &= \mathcal{A}^{(l)[j]}(\bar{u}_1^+, g_2^{h_2}, g_3^{h_3}, d_4^-; \bar{e}^+_5, \nu_6^-), \\
  \mathcal{A}^{(l)[j]}_{\Quark;h_2 h_3} &= \mathcal{A}^{(l)[j]}(\bar{u}_1^+, Q_2^{h_2}, \bar{Q}_3^{h_3}, d_4^-; \bar{e}_5^+, \nu_6^-),
\end{split}\end{align}
where we recall that we label all particles as outgoing. The numerical results we present are normalized
to the corresponding tree amplitudes to make the pole structure manifest.\\

\begin{table}[h]
  \newcolumntype{L}{>{\centering\arraybackslash}m{15ex}<{}}
  \renewcommand{\arraystretch}{1.3}
  \begin{adjustbox}{width=1\textwidth}
    \centering
    \begin{tabular}{l*{5}{L}}
      \toprule
       &   $\epsilon^{-2}$   &   $\epsilon^{-1}$   &   $\epsilon^{0}$   &   $\epsilon^{1}$   &   $\epsilon^{2}$   \\
      \midrule
      $\mathcal{A}^{(1)[0]}_{\gluon;++}$ & $-3.000000000$ & $-1.156228461$ $-6.283185307\ii$ & $114.1571538$ $-22.50315905\ii$ & $390.7700369$ $-27.70171948\ii$ & $772.8514956$ $-47.35817758\ii$ \\
      $\mathcal{A}^{(1)[1]}_{\gluon;++}$ & $0$ & $0$ & $-80.09777916$ $+22.21916010\ii$ & $-381.6882614$ $+105.8804960\ii$ & $-1061.591691$ $+294.4860144\ii$ \\ 
      \midrule                             
      $\mathcal{A}^{(1)[0]}_{\gluon;-+}$ & $-3.000000000$ & $-1.156228461$ $-6.283185307\ii$ & $9.229250985$ $-11.92991811\ii$ & $21.88019567$ $-11.15992632\ii$ & $28.67469046$ $+1.329575472\ii$ \\
      $\mathcal{A}^{(1)[1]}_{\gluon;-+}$ & $0$ & $0$ & $0$ & $0$ & $0$ \\ 
      \midrule                             
      $\mathcal{A}^{(1)[0]}_{\gluon;+-}$ & $-3.000000000$ & $-1.156228461$ $-6.283185307\ii$ & $10.84787275$ $-11.33482230\ii$ & $28.55563117$ $-7.684296706\ii$ & $42.82153823$ $+12.66461492\ii$ \\
      $\mathcal{A}^{(1)[1]}_{\gluon;+-}$ & $0$ & $0$ & $0$ & $0$ & $0$ \\ 
      \midrule                             
      $\mathcal{A}^{(1)[0]}_{\Quark;-+}$ & $-2.000000000$ & $2.109050494$ $-6.283185307\ii$ & $16.14776220$ $-12.06220529\ii$ & $33.34325904$ $-10.91964040\ii$ & $44.03129425$ $+3.915968598\ii$ \\
      $\mathcal{A}^{(1)[1]}_{\Quark;-+}$ & $0$ & $-0.6666666667$ & $-1.843519304$ & $-3.148759359$ & $-3.759742920$ \\ 
      \midrule                             
      $\mathcal{A}^{(1)[0]}_{\Quark;+-}$ & $-2.000000000$ & $2.109050494$ $-6.283185307\ii$ & $17.40843911$ $-12.56539566\ii$ & $39.85865914$ $-10.64655144\ii$ & $57.30017001$ $+11.51871433\ii$ \\
      $\mathcal{A}^{(1)[1]}_{\Quark;+-}$ & $0$ & $-0.6666666667$ & $-1.843519304$ & $-3.148759359$ & $-3.759742920$ \\
      \bottomrule
    \end{tabular}
  \end{adjustbox}
  \caption{Bare one-loop amplitudes of \cref{eq:amplitudesNum} evaluated at the phase-space point of 
  \cref{eq:psPointNum}, normalized by the corresponding tree amplitude. 
  The numerical results are rounded to fit the table.} 
  \label{tab:values-bare-amplitudes-1l}
\end{table}

\begin{table}[h]
  \newcolumntype{L}{>{\centering\arraybackslash}m{15ex}<{}}
  \renewcommand{\arraystretch}{1.3}
  \begin{adjustbox}{width=1\textwidth}
    \centering
    \begin{tabular}{l*{5}{L}}
      \toprule
       &   $\epsilon^{-4}$   &   $\epsilon^{-3}$   &   $\epsilon^{-2}$   &   $\epsilon^{-1}$   &   $\epsilon^{0}$   \\
      \midrule
      $\mathcal{A}^{(2)[0]}_{\gluon;++}$ & $4.500000000$ & $0.7186853827$ $+18.84955592\ii$ & $-366.7779225$ $+63.25510176\ii$ & $-1031.818957$ $-703.7101696\ii$ & $2229.756465$ $-4271.317939\ii$ \\
      $\mathcal{A}^{(2)[1]}_{\gluon;++}$ & $0$ & $0.5000000000$ & $241.5120803$ $-64.56308519\ii$ & $1008.875317$ $+259.9002941\ii$ & $-311.1773011$ $+3729.331393\ii$ \\
      $\mathcal{A}^{(2)[2]}_{\gluon;++}$ & $0$ & $0$ & $0$ & $53.39851944$ $-14.81277340\ii$ & $402.1206430$ $-111.5484479\ii$ \\ 
      \midrule                             
      $\mathcal{A}^{(2)[0]}_{\gluon;-+}$ & $4.500000000$ & $0.7186853827$ $+18.84955592\ii$ & $-51.99421404$ $+31.53537893\ii$ & $-122.1308176$ $-67.51062386\ii$ & $-13.35042281$ $-373.3014208\ii$ \\
      $\mathcal{A}^{(2)[1]}_{\gluon;-+}$ & $0$ & $0.5000000000$ & $1.218742820$ $+2.094395102\ii$ & $-4.764106945$ $+11.44393724\ii$ & $-39.16109323$ $+32.37695552\ii$ \\
      $\mathcal{A}^{(2)[2]}_{\gluon;-+}$ & $0$ & $0$ & $0$ & $0$ & $0$ \\ 
      \midrule                             
      $\mathcal{A}^{(2)[0]}_{\gluon;+-}$ & $4.500000000$ & $0.7186853827$ $+18.84955592\ii$ & $-56.85007933$ $+29.75009152\ii$ & $-134.3545770$ $-86.61366193\ii$ & $31.37211219$ $-446.6498136\ii$ \\
      $\mathcal{A}^{(2)[1]}_{\gluon;+-}$ & $0$ & $0.5000000000$ & $1.218742820$ $+2.094395102\ii$ & $-5.843188121$ $+11.04720671\ii$ & $-49.02848488$ $+27.67994158\ii$ \\
      $\mathcal{A}^{(2)[2]}_{\gluon;+-}$ & $0$ & $0$ & $0$ & $0$ & $0$ \\ 
      \midrule                             
      $\mathcal{A}^{(2)[0]}_{\Quark;-+}$ & $2.000000000$ & $-6.051434322$ $+12.56637061\ii$ & $-48.02138179$ $-0.6463175652\ii$ & $-45.51885333$ $-162.3399739\ii$ & $286.8120373$ $-570.6619745\ii$ \\
      $\mathcal{A}^{(2)[1]}_{\Quark;-+}$ & $0$ & $1.666666667$ & $0.9113214460$ $+6.283185307\ii$ & $-25.99983021$ $+31.15677229\ii$ & $-139.3729721$ $+83.16373240\ii$ \\
      $\mathcal{A}^{(2)[2]}_{\Quark;-+}$ & $0$ & $0$ & $0.4444444444$ & $2.458025738$ & $7.596909235$ \\ 
      \midrule                             
      $\mathcal{A}^{(2)[0]}_{\Quark;+-}$ & $2.000000000$ & $-6.051434322$ $+12.56637061\ii$ & $-50.54273561$ $+0.3600631747\ii$ & $-54.42997862$ $-173.7135037\ii$ & $349.9447856$ $-655.1503315\ii$ \\
      $\mathcal{A}^{(2)[1]}_{\Quark;+-}$ & $0$ & $1.666666667$ & $0.9113214460$ $+6.283185307\ii$ & $-27.68073275$ $+31.82769278\ii$ & $-154.3135540$ $+84.48223149\ii$ \\
      $\mathcal{A}^{(2)[2]}_{\Quark;+-}$ & $0$ & $0$ & $0.4444444444$ & $2.458025738$ & $7.596909235$ \\
      \bottomrule
    \end{tabular}
  \end{adjustbox}
  \caption{Bare two-loop amplitudes of \cref{eq:amplitudesNum} evaluated at the phase-space point of 
  \cref{eq:psPointNum}, normalized by the corresponding tree amplitude. 
  The numerical results are rounded to fit the table.} 
  \label{tab:values-bare-amplitudes-2l}
\end{table}

\begin{table}[h]
  \renewcommand{\arraystretch}{1.3}
    \centering
    \begin{tabular}{l*{2}{>{\centering\arraybackslash$}m{40ex}<{$}}}
      \toprule
        & l=1 & l=2 \\
      \midrule
      $\mathcal{R}^{(l)[0]}_{\gluon;++}$ & 103.3428710 - 16.56256597 \ii 	& 2630.789609 - 1047.648812 \ii \\
      $\mathcal{R}^{(l)[1]}_{\gluon;++}$ & -80.22358596 + 21.17196255 \ii 	& -2223.294883 + 1107.326463 \ii \\
      $\mathcal{R}^{(l)[2]}_{\gluon;++}$ & \text{--} 				& 179.9715930 + 40.91323686 \ii \\ 
      \midrule                                                                    
      $\mathcal{R}^{(l)[0]}_{\gluon;-+}$ & -1.585031823 - 5.989325032 \ii 	& 49.20099946 - 100.5754055 \ii \\
      $\mathcal{R}^{(l)[1]}_{\gluon;-+}$ & -0.1258067916 - 1.047197551 \ii 	& -15.30031141 + 26.62901308 \ii \\
      $\mathcal{R}^{(l)[2]}_{\gluon;-+}$ & \text{--} 				& -1.034903762 + 0.7918104613 \ii \\ 
      \midrule                                                                    
      $\mathcal{R}^{(l)[0]}_{\gluon;+-}$ & 0.03358994101 - 5.394229228 \ii 	& 76.72832903 - 103.5209963 \ii \\
      $\mathcal{R}^{(l)[1]}_{\gluon;+-}$ & -0.1258067916 - 1.047197551 \ii 	& -20.29786347 + 22.47920171 \ii \\
      $\mathcal{R}^{(l)[2]}_{\gluon;+-}$ & \text{--} 				& -1.034903762 + 0.7918104613 \ii \\ 
      \midrule                                                                    
      $\mathcal{R}^{(l)[0]}_{\Quark;-+}$ & 3.778746723 - 7.168809765 \ii 	& 109.4815548 - 140.5062051 \ii \\
      $\mathcal{R}^{(l)[1]}_{\Quark;-+}$ & -1.843519304 			& -45.62311230 + 41.31912332 \ii \\
      $\mathcal{R}^{(l)[2]}_{\Quark;-+}$ & \text{--} 				& 3.398563423 \\ 
      \midrule                                                                    
      $\mathcal{R}^{(l)[0]}_{\Quark;+-}$ & 5.039423632 - 7.672000135 \ii 	& 146.6740552 - 158.0359288 \ii \\
      $\mathcal{R}^{(l)[1]}_{\Quark;+-}$ & -1.843519304 			& -51.87649404 + 43.00174103 \ii \\
      $\mathcal{R}^{(l)[2]}_{\Quark;+-}$ & \text{--} 				& 3.398563423 \\
      \bottomrule
    \end{tabular}
  \caption{One- and two-loop finite remainders associated to the amplitudes of 
  \cref{eq:amplitudesNum} evaluated at the phase-space point of 
  \cref{eq:psPointNum}, normalized by the corresponding tree amplitude.
  The numerical results are rounded to fit the table.} 
  \label{tab:values-finite-remainders}
\end{table}

\FloatBarrier

\bibliographystyle{JHEP}
\bibliography{main}

\end{document}